# A Quadratically Regularized Functional Canonical Correlation Analysis for Identifying the Global Structure of Pleiotropy with NGS Data

Nan Lin,[1] Yun Zhu,[2] Ruzong Fan[3] and Momiao Xiong[1,*]

[1] Department of Biostatistics, School of Public Health, The University of Texas Health Science Center at Houston, Houston, TX 77225, USA.

[2] Department of Epidemiology, Tulane University School of Public Health and Tropical Medicine, New Orleans, LA 70112, USA.

[3] Biostatistics and Bioinformatics Branch (BBB), Division of Intramural Population Health Research (DIPHR), *Eunice Kennedy Shriver* National Institute of Child Health and Human Development, National Institutes of Health (NIH), Bethesda, MD 20892, USA.

Present Address: Department of Biostatistics, School of Public Health, The University of Texas Health Science Center at Houston, P.O. Box 20186, Houston, Texas 77225, (Phone): 713-500-9894, (Fax): 713-500-0900

*Correspondence: Momiao.Xiong@uth.tmc.edu

**Abstract**

Investigating the pleiotropic effects of genetic variants can increase statistical power, provide important information to achieve deep understanding of the complex genetic structures of disease, and offer powerful tools for designing effective treatments with fewer side effects. However, the current multiple phenotype association analysis paradigm lacks breadth (number of phenotypes and genetic variants jointly analyzed at the same time) and depth (hierarchical structure of phenotype and genotypes). A key issue for high dimensional pleiotropic analysis is to effectively extract informative internal representation and features from high dimensional genotype and phenotype data. To explore multiple levels of representations of genetic variants, learn their internal patterns involved in the disease development, and overcome critical barriers in advancing the development of novel statistical methods and computational algorithms for genetic pleiotropic analysis, we proposed a new framework referred to as a quadratically regularized functional CCA (QRFCCA) for association analysis which combines three approaches: (1) quadratically regularized matrix factorization, (2) functional data analysis and (3) canonical correlation analysis (CCA). Large-scale simulations show that the QRFCCA has a much higher power than that of the nine competing statistics while retaining the appropriate type 1 errors. To further evaluate performance, the QRFCCA and nine other statistics are applied to the whole genome sequencing dataset from the TwinsUK study. We identify a total of 79 genes with rare variants and 67 genes with common variants significantly associated with the 46 traits using QRFCCA. The results show that the QRFCCA substantially outperforms the nine other statistics.

**Introduction**

   As of December 18th, 2014, a catalog of published Genome-Wide Association Studies (GWAS) had reported significant association of 15,177 SNPs with more than 700 traits in 2,502 publications.[1] It is reported that more than 4.6% of the SNPs and 16.9% of the genes were significantly associated with more than one trait[2]. These results demonstrate that genetic pleiotropic effects, which refers to the effects of a genetic variant affecting multiple traits, play a crucial role in uncovering genetic structures of correlated phenotypes.[3-10] Most genetic analyses of quantitative traits have focused on a single trait association analysis, analyzing each phenotype independently.[11] Less attention has been paid to comprehensive analysis of pleiotropic effects.[12] However, multiple phenotypes are correlated due to shared genetic and environmental effects.[13] The integrative analysis of correlated phenotypes which tests the association of a genetic variant with multiple traits often increases the statistical power to identify genetic associations and increases the precision of genetic effect estimation.[13-16] It is increasingly recognized that the genetic effect can be detected only when the association of the genetic variant with the multiple traits are jointly tested.[17] It is also noted that directional pleiotropy widely exists. Changes of one trait may cause undesired changes of other traits. Investigation of pleiotropy provides a tool for designing the effective treatment with fewer side effects.

   Two types of approaches can be used for genetic pleiotropic analysis. One approach is to utilize summary statistics for estimating genetic correlations and testing association of genetic variants with multiple traits[17-21]. An alternative approach is to use individual genotypic information for association analysis of multiple correlated traits[22]. The focus of this paper is to use individual genotypes for pleiotropic analysis. Three major types of methods are commonly used to explore the association of genetic variants with multiple correlated phenotypes: multivariate techniques including multivariate linear models[15, 23-32], linear mixed models[16,32-34] and functional linear models[35], the combinations of univariate association measures for different phenotypes[36-39], and dimension reduction methods including principal component analysis (PCA),[14,40-43] and canonical correlation analysis.[44-48]

Statistical methods for testing the association of common variants with multiple traits have been well developed and successfully applied.[49] The methods for pleiotropic analysis of rare variants are still under development.[24,50] Next-generation sequencing and modern biosensing techniques have generated dozens of millions of SNPs and large numbers of clinical and intermediate phenotypes. The current multiple phenotype association analysis paradigm lacks breadth (the number of phenotypes and genetic variants jointly analyzed at a time) and depth (hierarchical structure of phenotype and genotypes). Most approaches perform analysis on the subsets of the full data space that are often missing, but now available. A key issue for high dimensional pleiotropic analysis is to effectively extract the informative internal representation and features from extremely high dimensional genotypic and phenotypic data. The statistical power of the methods that do not efficiently explore dimension reduction of both phenotype and genotype data will be limited. Despite their wide applications to the pleiotropic analysis, the current pleiotropic analysis methods share the same drawbacks. These methods, particularly multivariate analysis methods, either do not use data dimension reduction or ignore the rich linkage disequilibrium structure of genomic data when data dimension reduction is used. The most widely used methods for pleiotropic analysis are originally designed for analyzing a small number of phenotypes and common variant data. Due to the lack of efficient analytic platforms, the current pleiotropic analysis methods have not been applied to large-scale real genetic pleiotropic analysis with a large number of phenotypes and next-generation sequencing (NGS) data. To overcome these limitations and fully take the advantages of the genomic structure across a genomic region, we combine two approaches: (1) functional data analysis and (2) quadratically regularized CCA to develop a novel statistical method that is referred to as a quadratically regularized functional canonical correlation analysis (QRFCCA) for testing the association of genomic regions with multiple traits. The QRFCCA first transforms the high dimensional correlated discrete genotype data across the genes or genomic regions to a few regularized functional principal components in the low orthonormal eigenfunctional space by FPCA. Then, the QRFCCA will further utilize the quadratically recognized matrix factorization to project both the phenotype data and compressed genomic data by FPCA to low dimensional space with much fewer number of bases (components) than the traditional matrix factorization or PCA and changed distribution of eigenvalues in which the proportion of top eigenvalues substantially increases. The QRFCCA dramatically reduces the dimensions of both genotype and phenotype data while fully retaining the original genotypic and phenotypic information.

To evaluate the performance of the developed QRFCCA for association analysis of multiple phenotypes, we conduct large-scale simulations comparing QRFCCA to nine statistics: Sparse CCA (SCCA),[53] GAMuT[24], FCCA[51], kernel CCA (KCCA)[89], CCA, A Unified Score-Based Association Test (USAT)[54], PCA (applying to both phenotypes and genotypes), MANOVA (multivariate ANOVA applied to multiple phenotypes and SNPs), and minP (minimum of P-values for testing the association of single SNP with multiple phenotypes) and demonstrate that the QRFCCA has a much higher power than other competing statistics while retaining the correct type 1 error rates. Finally, the QRFCCA and nine other statistics are applied to the whole genome sequencing dataset from the TwinsUK study where 756 individuals with 33,746 genes and 46 traits in 13 major phenotype groups are included in the analysis. We find that the QRFCCA for pleiotropic analysis substantially outperforms the nine other statistics. A program for implementing the developed QRFCCA for association analysis of multiple phenotypes can be downloaded from our website http://www.sph.uth.tmc.edu/hgc/faculty/xiong/index.htm.

**Material and Methods**

To efficiently explore the organized genomic structure information in the data, the proposed QRFCCA for pleiotropic genetic analysis considers three levels of representation of genomic data within a gene or genomic region: (1) functional principal component representation, (2) matrix factorization and (3) quadratic regularization. In addition, there are two levels of representations of the unorganized phenotype data: (1) matrix factorization and (2) quadratic regularization. Genotype-phenotype association attempts to unravel the relationships between certain combinations of genetic variants from multiple loci and certain combinations of multiple phenotypes. The CCA measures the linear relationships between two multidimensional sets of variables and hence will be naturally used as a general framework for identifying the association between genotypes and phenotypes.

**Smooth Functional Principal Component Analysis**

For the self-contained, we briefly introduce the smooth functional principal component analysis (FPCA) for genetic variant data[55]. We first review the definition of genetic variant profiles. Let $t$ be the position of a genetic variant within a genomic region and $T$ be the length of the genomic region being considered. For convenience, we rescale the region from $[0,T]$ to $[0,1]$. We can view $t$ as a continuous variable in the interval $[0,1]$ because the density of genetic variants is high, We define the genotype of the $i$-th individual as

$$X_i(t) = \begin{cases} 2 & \text{MM} \\ 1 & \text{Mm} \\ 0 & \text{mm} \end{cases}, \qquad i = 1, \cdots, n \tag{1}$$

where M is an allele at the genomic position $t$ and $n$ is the number of sampled individuals.

To capture the variation of the genetic variant function, we define a linear combination of functional values:

$$f = \int_0^1 \beta(t) X(t) dt, \tag{2}$$

where $\beta(t)$ is a weight function and $X(t)$ is a centered genetic variant function defined in equation (1). The functional principal components can be obtained by choosing the weight function $\beta(t)$ to maximize the variance of $f$:

$$Var(f) = \int_0^1 \int_0^1 \beta(s) R(s,t) \beta(t) ds dt, \tag{3}$$

where $R(s,t)$ is the covariance function of the genetic variant function $X(t)$.

To improve the smoothness of the estimated functional principal component curves, we impose the roughness penalty on the functional principal component weight functions. The balance between the goodness-of-fit and the roughness of the function is controlled by a smoothing parameter $\mu$. The smoothed functional principal components (FPCs) can be obtained by solving the following integral equations:

$$\int_0^1 R(s,t) \beta(s) ds = \lambda [\beta(t) + \mu D^4 \beta(t)]. \tag{4}$$

Let $\beta_j(t)$, $j = 1, 2, \ldots$ be a set of FPCs. The genotype profile function $x_i(t)$ can be expanded in terms of orthogonal FPCs:

$$x_i(t) = \sum_{j=1}^{J} \xi_{ij} \beta_j(t),  \tag{5}$$

where $\xi_{ij}$ is the FPC score of the $i^{th}$ individual which can be estimated by

$$\xi_{ij} = <x_i(t), \beta_j(t)>_\mu , \ j = 1,2,...,k, \tag{6}$$

where $<.>_\mu$ is the inner product defined as

$$(f,g)_\mu = \int f(t)g(t)dt + \mu \int D^2 f(t) D^2 g(t) dt, \text{ where } D^2 f(t) = \frac{d^2 f(t)}{dt^2}.$$

**Matrix Factorization**

Consider a data matrix $A \in R^{n \times q}$ consisting of $n$ samples with $q$ features (variables). The data matrix $A$ represents the genotype data, the FPC scores or the phenotype data. The phenotype data include both continuous and discrete values. The $i^{th}$ row of $A$ is a vector of $q$ features for the $i^{th}$ sample, and the $j^{th}$ column of $A$ is a vector of the $j^{th}$ feature across the set of $n$ samples. Matrix factorization is used as a general framework to embed the genetic and phenotype data into the low dimensional vector space to reduce the data dimension and remove anomalous or noise data points[56]. To accomplish this, we first seek the best rank-$l$ approximation to the matrix $A$ by factorizing it into a product of two low rank matrices.

Let $G \in R^{n \times l}$ and $H \in R^{l \times q}$. Assume that the rank of $A$ is $r$. Therefore, $r \leq \min(n, q)$. Matrix factorization attempts to minimize the approximation error:

$$\min_{G,H} \ \|A - GH\|_F^2, \tag{7}$$

where $\|.\|_F$ denotes the Frobenius norm of a matrix.

Let $g_i \in R^{1 \times l}$ be the $i^{th}$ row of $G$, $h_j \in R^l$ be the $j^{th}$ column of $H$ and $A_{ij}$ be the value of the $j^{th}$ feature in the $i^{th}$ sample. Define $g_i h_j = (GH)_{ij}$ as an inner product. The objective function in problem (7) can be rewritten as

$$\sum_{i=1}^{n} \sum_{j=1}^{q} (A_{ij} - g_i h_j)^2 . \tag{8}$$

A solution to problem (7) can be found by truncating the singular value decomposition (SVD) of $A_{ij}$.[56] Let the SVD of $A$ be given by

$$A = U \Lambda V^T , \tag{9}$$

where $U = [u_1,...,u_r] \in R^{n \times r}$, $V = [v_1,...,v_r] \in R^{q \times r}$, $U^T U = I_{r \times r}, V^T V = I_{r \times r}$, and $\Lambda = \text{diag}(\lambda_1,...,\lambda_r) \in R^{r \times r}$ with $\lambda_1 \geq \lambda_2 \geq ... \geq \lambda_r > 0$. The columns of $U$ and $V$ are referred to as the left and right singular vectors of $A$, respectively, and $\lambda_1,...,\lambda_r$ are referred to as the singular values of $A$.

Substituting equation (9) into equation (7) gives

$$\min_{G,H} \quad \| U \Lambda V^T - GH \|_F^2 ,$$

which can be reduced to

$$\min_{G,H} \quad \| \Lambda - U^T GHV \|_F^2 . \tag{10}$$

Let $Z = U^T GHV$ and the rank of $Z$ be $l$. Then, we have

$$\| \Lambda - Z \|_F^2 = \sum_{i=1}^{n} \sum_{j=1, j \neq i}^{r} Z_{ij}^2 + \sum_{i}^{r} (Z_{ii} - \lambda_i)^2 . \tag{11}$$

To minimize $\| \Lambda - Z \|_F^2$, it must be $Z_{ij} = 0, i \neq j, Z_{ii} = 0, \forall i > l$ and $Z_{ii} = \lambda_i, i = 1,...,l$. In other words, the matrix which minimizes equation (11) should be

$$Z = \begin{bmatrix} \Lambda_l & 0 \\ 0 & 0 \end{bmatrix}, \tag{12}$$

where $\Lambda_l = \text{diag}(\lambda_1, ..., \lambda_l)$. The error of $l$-rank matrix approximation to $A$ is

$$\|\Lambda - Z\|_F^2 = \sum_{i=l+1}^{r} \lambda_i^2. \tag{13}$$

As a consequence,

$$U^T G H V = \begin{bmatrix} \Lambda_l & 0 \\ 0 & 0 \end{bmatrix} \text{ or}$$

$$GH = U \begin{bmatrix} \Lambda_l & 0 \\ 0 & 0 \end{bmatrix} V^T = U_l \Lambda_l V_l^T, \tag{14}$$

where $U_l = [u_1, ..., u_l]$ and $V_l = [v_1, ..., v_l]$.

Define $G = U_l \Lambda_l^{1/2}$ and $H = \Lambda_l^{1/2} V_l$. The matrix factorization of $A$ is then given by

$$A \approx GH. \tag{15}$$

**Canonical Correlation Analysis**

An alternative to multivariate linear regression analysis, the CCA is a popular analytic platform for genetic pleiotropic analysis. The goal of CCA is to seek optimal correlation between linear combinations of two sets of variables: the set of traits and the set of SNPs. The CCA measures the strength of association between the multiple SNPs and the traits. The pairs of linear combinations are called canonical variates and their correlations are called canonical correlations[57].

Consider a phenotype matrix $Y = [Y_1, ..., Y_k]$ with $k$ traits and genotype matrix $X = [X_1, ..., X_p]$ with $p$ SNPs or FPC scores. We assume that $p + k$ variables $Z = [X^T, Y^T]^T$ jointly have the covariance matrix

$$\Sigma_{zz} = \begin{bmatrix} \Sigma_{xx} & \Sigma_{xy} \\ \Sigma_{yx} & \Sigma_{yy} \end{bmatrix}.$$

We denote linear combinations of the matrices $X$ and $Y$ by

$$u = XA, \quad v = YB.  \tag{16}$$

In a matrix form, CCA can be formulated as

$$\begin{aligned} \max_{\theta, A} \quad & \text{Tr}(A^T \Sigma_{xy} B) \\ \text{s.t.} \quad & A^T \Sigma_{xx} A = I, \, B^T \Sigma_{yy} B = I. \end{aligned} \tag{17}$$

Let

$$R^2 = \Sigma_{YY}^{-1/2} \Sigma_{YX} \Sigma_{XX}^{-1} \Sigma_{XY} \Sigma_{YY}^{-1/2}, \tag{18}$$

$K = \Sigma_{XX}^{-1/2} \Sigma_{XY} \Sigma_{YY}^{-1/2}$ and the SVD of $K$ be

$$K = U \Lambda V^T, \tag{19}$$

where $\Lambda = \text{diag}(\lambda_1, ..., \lambda_q)$ and $q = \min(p, k)$ is the smaller number of variables in the two genotype-phenotype datasets. It is clear that

$$K^T K = R^2 = V \Lambda^2 V^T.$$

It is well known that the canonical vectors are

$$\begin{aligned} A &= \Sigma_{xx}^{-1/2} U, \\ B &= \Sigma_{YY}^{-1/2} V, \end{aligned} \tag{20}$$

and the vector of canonical correlations are

$$CC = [\lambda_1,...,\lambda_q]^T.  \tag{21}$$

A squared canonical correlation measures the proportion of variance linearly shared by the two sets of canonical variates derived from the input genotype-phenotype data sets.

Canonical correlations between the genotype and phenotypes measure the strength of their association. The CCA produces multiple canonical correlations. But we wish to use a single number to measure the association of the genetic variation with the multiple traits. We propose to use the summation of the square of the singular values as a measure to quantify the association of the genetic variation within a gene or genomic region with the multiple traits:

$$r = \sum_{i=1}^{q} \lambda_i^2 = \text{Tr}(\Lambda^2) = \text{Tr}(R^2). \tag{22}$$

To test the association of the genetic variation in a gene or genomic region is equivalent to test independence between the two genotype-phenotype datasets $X$ and $Y$ or to test the hypothesis that each variable in the set $X$ is uncorrelated with each variable in the set $Y$. The null hypothesis of no association of the genotype data $X$ with the phenotype dataset $Y$ can be formulated as

$$H_0 : \Sigma_{xy} = 0.$$

The likelihood ratio for testing $H_0 : \Sigma_{xy} = 0$ is

$$\Lambda_r = \frac{|\Sigma_{zz}|}{|\Sigma_{xx}||\Sigma_{yy}|} = \prod_{i=1}^{q}(1-\lambda_i^2), \tag{23}$$

which is equal to the Wilks' lambda $\Lambda$ defined in the multivariate linear regression model.

This demonstrates that testing for association using multivariate linear regression can be treated as special case of CCA[61].

We usually define the likelihood ratio test statistic for testing the association as:

$$T_{CCA} = -N\sum_{i=1}^{q} \log(1-\lambda_i^2). \tag{24}$$

For small $\lambda_i^2$, $T_{CCA}$ can be approximated by $N\sum_{i=1}^{q}\lambda_i^2 = Nr$, where $r$ is the measure of association of the genetic variation in the gene or genomic region with the multiple traits. The stronger the association, the higher the power that the test statistic can test the association.

Under the null hypothesis $H_0 : \Sigma_{xy} = 0$, $T_{CCA}$ is asymptotically distributed as a central $\chi^2_{pk}$. When sample size is large, Bartlett (1939) suggests using the following statistic for hypothesis testing:

$$T_{CCA} = -[N - \frac{(q+3)}{2}]\sum_{i=1}^{q} \log(1-\lambda_i^2). \tag{25}$$

**Quadratically Regularized Matrix Factorization and Canonical Correlation Analysis**

The power of the test statistics in CCA depends on the squared canonical correlations or eigenvalues of the matrix $R^2$. We wish to increase the power via changing distribution of the canonical correlations and data reduction. In matrix factorization, for fixed rank $l$, we want to approximate the matrix $A$ by the product of two factor matrices $G$ and $H$ as accurately as possible. However, the Frobenius norm of the matrices $G$ and $H$ may be large. We need to balance the approximation accuracy and the Frobenius norm of the factor matrices. Specifically, we add the Frobenius norm of the factor matrices to the objective in equation (7). The optimization problem (7) is now transformed to the quadratically regularized matrix factorization problem:

$$\min_{G,H} \quad F = \| A - GH \|_F^2 + \mu \| G \|_F^2 + \mu \| H \|_F^2. \tag{26}$$

From equation (15), the matrices $G$ and $H$ have the forms:

$$G = U_l \Lambda_l^{1/2} \text{ and } H = \Lambda_l^{1/2} V_l. \tag{27}$$

Substituting equation (27) into equation (26) gives

$$\min_{\Lambda_l} \quad F = \| \Lambda - \Lambda_l \|_F^2 + 2\mu \text{Tr}(\Lambda_l), \tag{28}$$

where $\Lambda = \text{diag}(\lambda_1,...,\lambda_k)$ and $\Lambda_l = \text{diag}(\tau_1,...,\tau_l,0,...,0)$.

Expanding the Frobenius norm of the matrices, problem (28) can be further reduced to

$$\min_{\tau} \quad F = \sum_{i=1}^{l} (\lambda_i - \tau_i)^2 + \sum_{i=l+1}^{r} \lambda_i^2 + 2\mu \sum_{i=1}^{l} \tau_i. \tag{29}$$

To minimize $F$, we set

$$\frac{\partial F}{\partial \tau_j} = -2(\lambda_j - \tau_j) + 2\mu = 0. \tag{30}$$

Solving equation (30) for $\tau_j$ yields

$$\tau_j = \lambda_j - \mu. \tag{31}$$

To ensure that the factor matrices are non-negative, $\tau_j$ must be non-negative. Thus, we have the solution:

$$\tau_j = (\lambda_j - \mu)_+, \tag{32}$$

where $(a)_+ = \max(a,0)$.

Define the matrix $\Lambda_l$ as

$$\Lambda_l = \text{diag}((\lambda_1 - \mu)_+, \ldots, (\lambda_l - \mu)_+).$$

Then, factor matrices $G = U_l \Lambda_l^{1/2}$ and $H = \Lambda_l^{1/2} V_l$ are the solution to the minimization problem (26). We use truncation of the SVD to keep only the top $l$ singular values and soft-thresholding on the singular values to change distribution of the singular values. When $\mu$ increases beyond some singular values $\lambda_m$, $l - m + 1$ singular values of $GH$ will disappear. Analytically, we can easily show that

$$\frac{\lambda_1 - \mu}{\lambda_1 - \mu + \lambda_2 - \mu + \ldots + \lambda_l - \mu} > \frac{\lambda_1}{\lambda_1 + \lambda_2 + \ldots + \lambda_l}.$$

In other words, increasing $\mu$ will move up the proportion of the first singular value in the total of singular values. The phenotype data consist of 756 samples with 46 traits. The initial largest singular value and total singular values of the phenotype data are 73.97 and 1030.14 respectively. Figure S1 shows that the proportion of the first singular value in the total of singular values is an increasing function of threshold $\mu$. This clearly demonstrates that adding quadratic regularization results in changing the distribution of the singular values of the factor matrices. Therefore, we can expect that regularized matrix factorization for data reduction will increase the power to detect association of the genetic variation with the traits.

Quadratically regularized matrix factorization and CCA have broad applications. There are a number of ways to use the quadratically regularized matrix factorization for data reduction in the association analysis which are briefly summarized as follows.

(1) Continuous phenotypes and NGS genotype data (dimension reduced genotype data using FPCA-gene-based association study).

We can first apply the quadratically regularized matric factorization to both multiple phenotype data and FPC score data in a gene or genomic region and then use CCA for association analysis of multiple traits. This analysis will be referred to as quadratically regularized FCCA (QRFCCA) for multiple trait association analysis. If only a single trait is considered, the quadratically regularized matrix factorization is only applied to the FPC score in the gene.

(2) Continuous phenotypes and multiple SNPs.

Quadratically regularized matrix factorization is first applied to both multiple phenotypes and SNPs for data reduction and then CCA is used for multiple trait association analysis. This procedure is referred to as QRMCCA.

(3) Continuous phenotypes and a single SNP.

Quadratically regularized matrix factorization is first applied to multiple phenotypes and then CCA is used for multiple trait association analysis. This procedure is referred to as QRSCCA.

(4) Both functional phenotype such as RNA-seq data and functional genotype data such as NGS data. FPCA is first applied to both functional phenotype and genotype data to obtain FPC scores for both phenotypes and genotypes. Quadratically regularized matrix factorization is applied to FPC scores of both functional phenotype and genotype data. Finally, CCA is used for multiple trait association analysis. This procedure is referred to as QRBFCCA.

Quadratically recognized matrix factorization can also be applied to the $K$ or $R^2$ matrix in the CCA. The test statistics then use the singular values of the reduced $K$ or $R^2$ matrix to test association of genetic variation with a trait.

**Connection between CCA and Maximum Heritability Analysis**

Next we study the relationships between the CCA and the traditional quantitative trait analysis. For simplicity, we assume that the genetic variation is the only major contribution to the phenotypic variation and we will not consider the covariates. For the single trait, in Supplemental note A, we show that the narrow heritability can be expressed as the proportion of the phenotype variation explained by the genetic variation or the squared correlation between the genotype and phenotype.

Now we consider the multiple traits. We assume that the multivariate linear model for the multiple traits is given by

$$Y = XB + \varepsilon, \tag{33}$$

where $X$ is a genotype matrix and $B = [b_1,...,b_k]$ is the regression coefficient matrix or genetic effect matrix.

The genetic effect matrix can be estimated by

$$\hat{B} = \Sigma_{xx}^{-1}\Sigma_{xy}. \tag{34}$$

The covariance of the genetic additive effects is defined as (Supplemental note A, N7)

$$\Sigma_A = \Sigma_{yx}\Sigma_{xx}^{-1}\Sigma_{xy}. \tag{35}$$

The concept of heritability for a single trait is well developed. Recently, there has been increasing attempts to extend heritability from a single trait to multiple traits. However, the focus is to consider the heritability of a linear combination of multiple traits[50,59,60]. Here, we consider a heritability matrix that is defined as

$$h_m^2 = \Sigma_{yy}^{-1/2}\Sigma_{yx}\Sigma_{xx}^{-1}\Sigma_{xy}\Sigma_{yy}^{-1/2}, \tag{36}$$

which is equal to the matrix $R^2$ in the CCA (equation 18).

Let $K = \Sigma_{xx}^{-1/2}\Sigma_{xy}\Sigma_{yy}^{-1/2}$ and SVD of $K$ be

$$\Sigma_{xx}^{-1/2}\Sigma_{xy}\Sigma_{yy}^{-1/2} = U_K \Lambda_K V_K^T. \tag{37}$$

In Supplemental note A, we show that the maximum heritability analysis is equivalent to CCA. The optimal linear combination coefficients are given by

$$\begin{aligned} a &= \Sigma_{xx}^{-1/2} u_k \\ b &= \Sigma_{yy}^{-1/2} v_k, \end{aligned} \tag{38}$$

where $u_k$ and $v_k$ are the left and right singular vectors of the matrix $K$, respectively.

In Supplemental note A, we show that the genetic additive effect and genetic additive variance are, respectively, given by (N19)

$$\begin{aligned} \hat{\alpha} &= u_K^T K v_K = \tau_K \\ \sigma_{Al}^2 &= \hat{\alpha}^2 = \tau_K^2 = b^T \Sigma_{yx}\Sigma_{xx}^{-1}\Sigma_{xy} b, \end{aligned} \tag{39}$$

where $\tau_K$ is the singular value of the matrix $K$ and $\tau_K^2$ is the eigenvalue of the matrix

$$K^T K = \Sigma_{yy}^{-1/2}\Sigma_{yx}\Sigma_{xx}^{-1}\Sigma_{xy}\Sigma_{yy}^{-1/2} = R^2. \tag{40}$$

The narrow-sense heritability is

$$h_l^2 = \tau_K^2, \tag{41}$$

which is the square of canonical correlation. This shows that the maximum heritability analysis is equivalent to CCA.

**Relationship among CCA, Kernel CCA, Functional CCA, Cross-covariance operator, Dependence Measure and Other Association Tests**

In supplemental note B, we use reproducing kernel Hilbert spaces (RKHS) as a general framework and the covariance operator as a general tool for unifying CCA, kernel CCA, functional CCA and other association analyses including GAMuT. Many multivariate and functional statistical methods such as regression, CCA, kernel regression, kernel CCA, functional regression and functional CCA can be used to test the association of genetic variants with the phenotypes. In the past decades, reproducing kernel Hilbert spaces (RKHS) have emerged as a general framework for various statistical and machine learning methods. In supplemental note B, we use RKHS as a unified framework for association tests to reveal the relationships among various multivariate and functional association tests.

Covariance is a key measure for assessing association. In Supplemental Note B, we extend covariance between two random variables in the original probability space (N24):

$$\tilde{\Sigma}_{XY} = E[XY^T] \tag{42}$$

to the feature space (N25)

$$\tilde{\Sigma}_{XY} = E[\phi(X)\psi^T(Y)], \tag{43}$$

where $\phi(X)$ and $\psi(Y)$ are feature maps and $\tilde{\Sigma}_{XY}$ is referred to as a covariance operator. We then show that

$$f^T \tilde{\Sigma}_{XY} g = E[f(X)(g(Y)], \tag{44}$$

where $f$ and $g$ are vectors in the feature space. We show that (N33)

$$f^T \Sigma_{XY} g = \frac{1}{m} \alpha^T \tilde{K}_x \tilde{K}_y \beta, \tag{45}$$

where $\Sigma_{XY} = \tilde{\Sigma}_{XY} - \mu_X \mu_Y^T$, $K_x = (K(X_i, X_j))_{m \times m}$, $K_y = (K(Y_i, Y_j))_{m \times m}$, $G = I_m - \frac{1}{m} \mathbf{1}_m$, $\tilde{K}_x = G K_x G$ and $\tilde{K}_y = G K_y G$.

Solving the optimization problem (N34) leads to the dependence measure (N39):

$$\frac{1}{m^2} \text{Trace } (\tilde{K}_{xx} \tilde{K}_{yy}), \tag{46}$$

which is the basis for the GAMuT test.[24] In Supplemental note B, we show that the KCCA is quite similar to the kernel independent test and if the constraints in (N40) in the KCCA are replaced by $Var(U) = \alpha^T \tilde{K}_x \alpha = 1$ and $var(V) = \beta^T \tilde{K}_y \beta = 1$, then the association measure in the KCCA is equal to

$$\frac{1}{m^2} \text{Tr}(\Lambda^2) = \frac{1}{m^2} \text{Trace } (\tilde{K}_{xx} \tilde{K}_{yy}),$$

which is exactly equal to the dependence measure.

Finally, we consider the FCCA. To unify multivariate association tests and functional association tests, we use RKHS as a general framework for formulation of the functional CCA.[1]

Consider two index sets $E_1$ and $E_2$, and two stochastic processes: $\{X(t), t \in E_1\}$ and $\{Y(s), s \in E_2\}$ with mean zero $E[X(t)] = E(Y(s)) = 0$ for all $t, s$, auto covariance functions $R_X(s_1, s_2) = \text{cov}(X(s_1), X(s_2))$, $R_Y(t_1, t_2) = \text{cov}(Y(t_1), Y(t_2))$ and cross covariance functions $R_{XY}(s,t) = \text{cov}(X(s), Y(t))$.

Let $L_X^2$ and $L_Y^2$ be the Hilbert spaces spanned by the $X$ and $Y$ processes. Let $\mathcal{H}(R_X)$ be RKHS generated by the covariance kernel $R_X$ and $\mathcal{H}(R_Y)$ be RKHS generated by the covariance kernel $R_Y$. The canonical correlation can be defined in terms of both $L_X^2$ and $L_Y^2$, and $\mathcal{H}(R_X)$ and $\mathcal{H}(R_Y)$:

$$\begin{aligned} & \max_{\substack{\zeta \in L_X^2, \upsilon \in L_Y^2 \\ \text{var}(\zeta)=1, \text{var}(\upsilon)=1}} \text{cov}^2(\zeta, \upsilon) \\ &= \max_{\substack{\alpha \in \mathcal{H}(R_X), \beta \in \mathcal{H}(R_Y) \\ \|\alpha\|_{\mathcal{H}(R_X)}^2 = 1, \|\beta\|_{\mathcal{H}(R_Y)}^2 = 1}} \text{cov}^2(\Psi_X(\alpha), \Psi_Y(\beta)), \end{aligned} \quad (47)$$

where $\Psi_X(\alpha)$ and $\Psi_Y(\beta)$ are the congruence mapping from $\mathcal{H}(R_X)$ to $L_X^2$ and $\mathcal{H}(R_Y)$ to $L_Y^2$, respectively.

We show that formulation of FCCA in equation (N65) is equivalent to the formulation of the FCCA in the $L_X^2, L_Y^2$ and $\mathcal{H}(R_X), \mathcal{H}(R_Y)$ as expressed in equation (47). Therefore, the FCCA can be defined as (N65)

$$\begin{aligned} \max_{\alpha, \beta} \quad & \alpha^T \Lambda_{XY} \beta \\ \text{s.t.} \quad & \alpha^T \Lambda_{XX} \alpha = 1, \beta^T \Lambda_{YY} \beta = 1, \end{aligned} \quad (48)$$

where

$$\Lambda_{XY} = \begin{bmatrix} \gamma_{11} & \cdots & \gamma_{1q} \\ \vdots & \vdots & \vdots \\ \gamma_{p1} & \cdots & \gamma_{pq} \end{bmatrix}, \Lambda_{XX} = \begin{bmatrix} \lambda_1 & \cdots & 0 \\ \vdots & \vdots & \vdots \\ 0 & \cdots & \lambda_p \end{bmatrix}, \Lambda_{YY} = \begin{bmatrix} \mu_1 & \cdots & 0 \\ \vdots & \vdots & \vdots \\ 0 & \cdots & \mu_q \end{bmatrix}, \alpha = \begin{bmatrix} \alpha_1 \\ \vdots \\ \alpha_p \end{bmatrix}, \beta = \begin{bmatrix} \beta_1 \\ \vdots \\ \beta_q \end{bmatrix}.$$

We show that the association measure $r$ of the FCCA is given by

$$r = \mathrm{Tr}(R^2) = \sum_{i=1}^{p} \sum_{j=1}^{q} \frac{\gamma_{ij}^2}{\lambda_i \mu_j}. \tag{49}$$

If constraints $\alpha^T \Lambda_{XX} \alpha = 1, \beta^T \Lambda_{YY} \beta = 1$ in equation (38) is replaced by $\alpha^T \alpha = 1$ and $\beta^T \beta = 1$, i.e., the optimization problem (48) is reduced to

$$\begin{aligned} \max_{\alpha, \beta} \quad & \alpha^T \Lambda_{XY} \beta \\ \text{s.t.} \quad & \alpha^T \alpha = 1, \beta^T \beta = 1, \end{aligned} \tag{50}$$

then equation (49) is reduced to

$$r = \mathrm{Tr}(R^2) = \sum_{i=1}^{p} \sum_{j=1}^{q} \gamma_{ij}^2. \tag{51}$$

Suppose that the FPC scores form a feature space. In supplemental note B, we define the feature maps from the original functional data to the FPC score feature space. We show that the dependence measure in the FPC score-based kernel analysis is asymptotically equal to the association measure of the FCCA. This implies that the FCCA is a specific kernel analysis that uses the FPC score to define the kernels instead of directly using the genotype data to define the kernels.

**Results**

**Null Distribution of Test Statistics**

To examine the null distribution of test statistics for association analysis of multiple traits, we performed a series of simulation studies to compare their empirical levels with the nominal ones. We calculated the type I error rates for rare alleles, and both rare and common alleles. We first assumed the model for multiple traits:

$Y_i = \mu + \varepsilon_i$, $i = 1,...,n,$

where $Y_i = [y_{i1},..., y_{ik}]$ $k$ is the number of traits, $\mu$ is an overall mean, and $\varepsilon_i$ is distributed as $N(0, \Sigma)$, where $\Sigma$ is a $k \times k$ residual correlation matrix. We similarly model the correlation matrix as in Broadaway et al[24]. We also consider three scenarios of low residual correlation among phenotypes with pair-wise correlation selected from a uniform (0.1, 0.2) distribution, moderate residual correlation with pair-wise correlation selected from a uniform (0.2, 0.4) distribution, and high residual correlation with pair-wise correlation selected from a uniform (0.4, 0.7) distribution.

We randomly generated 1,000,000 haplotypes with gene *C16orf62* from 659 samples of European origin in The 1000 Genome Project. 1,000 SNPs with 600 rare variants (frequencies ranging from 0.0005 to 0.01) and 400 common variants (frequencies larger than 0.01) were randomly selected from C16orf62 gene. The number of sampled individuals for type 1 error simulations from populations of 500,000 individuals ranged from 500 to 2,000. A total of 10,000 simulations were repeated. The type 1 error rates were estimated as the proportion of the datasets under the null distribution in which the P-values were less than or equal to the significance level.

Tables 1 and 2 summarized the type 1 error rates of the ten statistics: QRFCCA, Sparse CCA (SCCA)[53], GAMuT, FCCA, Kernel CCA (KCCA), CCA, A Unified Score-Based Association Test (USAT)[54], PCA (applying to both phenotypes and genotypes), MANOVA (multivariate

ANOVA applied to multiple phenotypes and multiple SNS) and minP (minimum of P-values for testing the association of single SNP with multiple phenotypes) for testing the association of rare variants, and both rare and common variants, within a genomic region with 15 high correlated traits, respectively, at the nominal levels α=0.05, α=0.01 and α=0.001. Tables S1-S16 showed type 1 error rates of the ten statistics for testing the association of rare variants, and both rare and common variants with 5, 10 and 15 traits under three scenarios: low, moderate and high correlations. These tables showed that the estimated type 1 error rates of the QRFCCA across a range of assumptions were not appreciably different from the nominal levels $\alpha = 0.05$, $\alpha = 0.01$ and $\alpha = 0.001$. We also observed that the type 1 error rates of other nine statistics, in most scenarios, were appropriate.

**Power Evaluation**

To evaluate the performance of the QRFCCA in association analysis, we used simulated data to estimate power of ten statistics for testing the association of a gene or a genomic region with the traits. We simulated 5, 10 and 15 traits with low, moderate and high correlations. An additive genetic model was used to summarize all genetic effects of causal variants in the gene or genomic region.

For each individual, 5, 10, 15 quantitative traits were simulated by the summation of genetic effect and the residual correlation between the traits. Let $h_k^2$ be the narrow heritability of the $k^{th}$ trait. Assume that each SNP had a 2% chance to be associated with a trait and its genetic effect on the $k^{th}$ trait was equal to the $\sqrt{\dfrac{h_k^2}{MAF}}$ multiplied by the number of minor alleles where $MAF$ denoted the frequency of the minor allele. This indicates that the genetic effect of causal variants was inversely proportional to its minor allele frequency. Therefore, we assumed that the very rare variants had larger genetic effects than less rare variants.

We did not assume that the gene of interest was associated with all traits. In fact, we assumed that the gene of interest was truly associated with three of five assessing traits, six of ten assessing traits and eight of 15 assessing traits. The residual correlation was simulated from a multivariate distribution with mean zero and covariance matrix

$$\begin{bmatrix} 1-h_1^2 & r_{12}\sqrt{(1-h_1^2)(1-h_2^2)} & \cdots & r_{1K}\sqrt{(1-h_1^2)(1-h_K^2)} \\ r_{12}\sqrt{(1-h_1^2)(1-h_2^2)} & 1-h_2^2 & \cdots & r_{2K}\sqrt{(1-h_2^2)(1-h_K^2)} \\ \vdots & \vdots & \vdots & \vdots \\ r_{K1}\sqrt{(1-h_1^2)(1-h_K^2)} & r_{K2}\sqrt{(1-h_2^2)(1-h_K^2)} & \cdots & 1-h_K^2 \end{bmatrix},$$

where the correlation between traits $r_{ij}$ was randomly generated with uniform distribution: low correlation $[0.1-0.2]$, moderate correlation $[0.2-0.4]$ and high correlation $[0.4-0.7]$. In summary, the genetic model for power evaluation is given by

$$[y_1 \ \cdots \ y_K] = [x_1 \ \cdots \ x_q] \times \left[ \begin{pmatrix} \alpha_{11} & \cdots & \alpha_{K1} \\ \vdots & \ddots & \vdots \\ \alpha_{1q} & \cdots & \alpha_{Kq} \end{pmatrix} \circ \begin{pmatrix} b_{11} & \cdots & b_{K1} \\ \vdots & \ddots & \vdots \\ b_{1q} & \cdots & b_{Kq} \end{pmatrix} \right] \times \begin{pmatrix} t_1 & \cdots & 0 \\ \vdots & \ddots & \vdots \\ 0 & \cdots & t_K \end{pmatrix}$$

$$\times \begin{pmatrix} h_1^2 & \cdots & 0 \\ \vdots & \ddots & \vdots \\ 0 & \cdots & h_K^2 \end{pmatrix} + [\varepsilon_1 \ \cdots \ \varepsilon_K],$$

where $y_i$ represented phenotypes, $x_j$ was an indicator variable for coding the genotype of the $j^{th}$ SNP in the gene, taking values 0, 1, 2 to represent the number of minor alleles at the SNP. $\alpha_{ij}$ denoted the genetic effect that followed a normal distribution with $N(0, \sqrt{\frac{h_i^2}{MAF_j}})$, where $h_i^2$ denoted the narrow heritability of the $i^{th}$ trait, $MAF_j$ denoted the frequency of the minor allele at the $j^{th}$ SNP, $b_{ij}$ in the matrix represented the probability of the $j^{th}$ SNP being the causal variant for the $i^{th}$ trait and followed a binomial distribution $B(1,0.02)$. Notation $\circ$ denoted an element-wise matrices multiplication, $t_i \sim B(1,0.6)$ represented the probability of the gene being tested contributing the genetic effect to the $i^{th}$ trait and $h_i^2$ denoted the heritability of the $i^{th}$ trait and followed a uniform distribution $U(0.005,0.015)$, $\varepsilon_i, i = 1,...,K$ denoted residuals and followed a multivariate normal distribution as defined above.

The genotype data in type 1 error calculations were also used for power evaluation. A total of 10,000 simulations were repeated for the power calculations.

We first compared the power of QRFCCA with nine other competing statistics that are described in type 1 error rate calculations for testing the association of rare variants with multiple continuous traits. Power was estimated as a function of sample sizes. Figures 1-3 plotted the power of the curves as a function of sample sizes of the ten statistics for collectively testing the association of all rare variants in the gene with 10 low, moderately and highly correlated traits, respectively, at the significance level $\alpha = 0.05$. The power curves of ten statistics for testing the association of the gene including only rare variants with 5 and 15 low, moderately and highly correlated traits, respectively, were plotted in Figures S2-S4, Figures S5-S7. We observed several remarkable features. First, we clearly observed that the QRFCCA had highest power among all ten statistics, followed by SCCA for all scenarios considered. Second, in general, the power of the FCCA was higher than that of the KCCA and the GAMuT, but their differences were very small. Third, we often observed that the functional data analysis-based and kernel-based methods had higher power than the classical multivariate methods.

Next we investigated whether the power pattern of the ten statistics for testing the association of the gene with only rare variants would still hold when testing the association of the gene with both rare and common variants. Figures 4-6 presented the power curves of ten statistics for testing the association of the gene including both rare and common variants with 10 low, moderately and highly correlated traits at the significance level $\alpha = 0.05$, respectively. Figures S8-S10, and S11-S13 showed the power curves of the ten statistics for testing the association of the gene including both rare and common variants with 5 low, moderately, and highly correlated traits, and 15 low, moderately, and highly correlated traits, respectively. We observed the similar power pattern of the tests for both rare and common variants as that for only rare variants. These figures demonstrated that the QRFCCA substantially outperformed the nine other statistics and the difference in power between the QRFCCA and other statistics for the both rare and common variants was much larger than that for the rare variants only. This demonstrated that the regularization in singular vectors plays a more important role in association analysis of both rare and common variants than that in association analysis of only rare variants.

 **Application to Real Data Examples**

Investigation of the contribution of the entire allelic spectrum of genetic variation to multiple traits are still at its infancy. The systematic searching for both common and rare variants associated with large number of traits is essential for unraveling the genetic architecture of complex diseases. To further evaluate the performance, the QRFCCA and nine other statistics were applied to the UK-10K dataset. The UK10K Cohorts project used a low read depth whole-genome sequencing (WGS) to assess the contribution of the genetic variants to the sixty-four different traits.[62] However, missing phenotypes were found in many individuals. To ensure no missing phenotypes in individuals, we included 765 individuals with 2,240,049 SNPs in 33,746 genes, and shared 46 traits in 13 major phenotypic groups which covered a wide range of traits (Table S17) in the analysis. We took the rank-based inverse normal transformation of the phenotypes[63] as trait values.

We first studied the association of genes with only rare variants (MAF $\leq 0.01$). The total number of genes with only rare variants tested for association was 33,746. A p-value for declaring significant interaction after applying the Bonferroni correction for multiple tests was $1.48 \times 10^{-6}$. To examine the behavior of the test statistics, we plotted the QQ plot of the QRFCCA with one FPC, FCCA with one FPC, PCA and GAMuT using a linear kernel in Figure 7 and the QQ plot of the KCCA, SCCA, CCA, MANOVA and USAT in Figure S14. The QQ plots showed that the false positive rate of the QRFCCA and FCCA for testing the association of the gene with 46 traits in some degree was controlled. However, the behavior of the QQ plot of KCCA and SCCA was weird.

The total number of genes significantly associated with the 46 traits using QFCCA, FCCA, PCA, SCCA, KCCA, GAMuT, CCA, USAT and MANOVA, were 79, 59, 14, 8, 3, 0, 0, 0 and 0, respectively (Table 3, Table S18). We observed that the list of 79 significant genes identified by QFCCA included all 59 significant genes using FCCA, all 8 significant genes using SCCA, all 3 significant genes using KCCA and 8 significant genes using PCA. The Manhattan plot showing genome-wide p-values of association with 46 traits calculated using QRFCCA is presented in Figure 8.

To further assess the performance of the QFCCA and GAMuT, we presented Tables S19 and S20. Table S19 summarized the top ten genes ranked using GAMuT where p-values calculated by both GAMuT and QRFCCA were also listed. None of ten genes reached genome-wide significance levels by the GAMuT. However, we noticed that 7 of top ten gens ranked by GAMuT were significantly associated with the 46 traits identified by QFCCA. Although we observed that the p-value of *RNU6-1229P* calculated by GAMuT was smaller than that calculated by QRFCCA, we did not find any significant SNPs within *RNU6-1229P* (Figure S15). This may imply that association was spurious. In Table S20, we listed p-values of all SNPs within gene *ADAM19*. We observed that QRFCCA identified significance *of ADAM19* with a p-value less than $6.07 \times 10^{-11}$ and at least 4 SNPs in *ADAM19* had very small p-values and two additional SNPs had p-values that were close to the threshold p-value of genome-wide significance (Figure S15). However, the GAMuT missed to identify significance of *ADAM19*.

To characterize the pleotropic pattern, we presented the heat map showing the pattern of cross phenotype association of genes with rare variants only and the most important pleiotropic effects of the genes (Figure 9). Table S21 summarizes the number of traits which a single gene was associated with (p-value $\leq 0.05$). In Table S21, we also listed the p-values for testing the association of the gene with all 46 traits. All p-values in Figure 9 and Table S21 were calculated using QRFCCA. We observed two remarkable features. First, we observed that 5 genes were significantly associated with 3 traits, 10 genes were significantly associated with 2 traits, and 39 genes were significantly associated with one trait at the genome-wide significance level after Bonferroni correction. The remaining 25 genes did not reach the genome-wide significance with any trait. However, we observed that these genes still showed mild association with multiple traits. Second, we observed that multiple genes were significantly associated with single phenotype (Table S22). For example, 345 genes were significantly associated with creatinine, 108 genes with HOMA-IR, 20 genes with HOMA-B, 72 genes with HsCRP, 21 genes with glucose, 15 genes with insulin, 14 genes with GGT, 11 genes with triglycerides and 11 genes with VLDL.

Some results can be confirmed in the literature. Throughout this section, all p-values were calculated using QRFCCA. *PNOC* which was associated with the 46 traits (p-value $\leq 1.63 \times 10^{-12}$), HOMA-IR (p-value $\leq 1.2 \times 10^{-7}$) and triglycerides (p-value $\leq 9.15 \times 10^{-9}$) was reported to be associated with insulin resistance[64] and triglycerides.[65] *MIR409* which was associated with the 46 traits (p-value $\leq 2.77 \times 10^{-10}$), weight (p-value $\leq 1.65 \times 10^{-6}$), Hip (p-value $\leq 7.19 \times 10^{-7}$) and total lean mass (p-value $\leq 2.15 \times 10^{-6}$) was used as a weight loss biomarker.[66] *GAPDH* which showed an association with the 46 traits (p-value $\leq 2.96 \times 10^{-8}$) and specifically with creatinine (p-value $\leq 3.45 \times 10^{-8}$) was reported to be associated with creatinine.[67] *MLN* which demonstrated association with the 46 traits (p-value $\leq 9.47 \times 10^{-8}$) and showed a strong association with glucose (p-value $\leq 6.24 \times 10^{-11}$) played an important role in controlling the rise rate of glucose level.[68] *JUN*, which was associated with the 46 traits (p-value $\leq 1.25 \times 10^{-7}$) and showed a strong association with creatinine (p-value $\leq 3.43 \times 10^{-20}$), was reported to be correlated with the serum creatinine level.[69] *COMP* which showed significance with the 46 traits p-value $\leq 4.83 \times 10^{-7}$) and a strong association with specific trait creatinine (p-value $\leq 9.58 \times 10^{-15}$) was also reported an association with creatinine.[70]

Next we studied the association of genes with only common variants (MAF $\geq 0.05$). The total number of genes with only common variants tested for association was 33,166. The p-value to declare the significant association after applying Bonferroni correction for multiple tests was $1.51 \times 10^{-6}$. To examine the behavior of the test statistics, we plotted the QQ plot of the QRFCCA, FCCA, PCA and GAMuT using a linear kernel in Figure 10 and the QQ plot of the KCCA, SCCA, CCA, MANOVA and USAT in Figure S16. Similar to rare variants, the QQ plots for common variants showed that the false positive rate of the QRFCCA and FCCA for testing the association of the gene with 46 traits in some degree was controlled. However, the behavior of the QQ plot of KCCA and SCCA was weird.

The total number of genes significantly associated with the 46 trait using QFCCA, FCCA, PCA, SCCA, KCCA, GAMuT, CCA, USAT and MANOVA, were 67, 55, 0, 0, 31, 0, 0, 0 and 0, respectively (Table 4, Table S23). Similar to rare variants, we observed that a list of 67 significant genes identified by QFCCA included all the 55 significant genes identified using FCCA. However, unlike rare variants, only one significant gene was shared between the QRFCCA and KCCA. To assess whether the QRFCCA for testing the association of genes including common variants only with multiple traits was appropriate or not, we presented Figure S17 which shows the p-values of all SNPs within gene *REG1B*. We observed in Figure S17 that more than 11 SNPs in *REG1B* with p-values $\leq 0.0001$ jointly made contributions to the strong association of *REG1B* with the 46 traits with p-value $\leq 1.65 \times 10^{-116}$. The QRFCCA can catch the features of genetic variation. In Figure S17 we listed the p-values of all SNPs within the gene *XXbac-BPG154L12.4*. From Figure S17, we also observed that although the GAMuT treated *XXbac-BPG154L12.4* as associated with the 46 traits (p-value $\leq 2.55 \times 10^{-6}$), none of SNPs were even weakly associated with the 46 traits. We should point out that the QRFCCA did not find any, even mild association (p-value $\leq 0.3175$). The Manhattan plot showing genome-wide p-values of association of genes consisting of only common variants with the 46 traits calculated using QRFCCA is presented in Figure 11.

To unravel the genetic pleotropic structure of common variants, we presented the gene/phenotype association heat map that demonstrated the most important pleiotropic relations between a single gene and multiple traits (Figure 12) and summarized the number of traits a single gene affected in Table S24. All p-values in Figure 12 and Table S24 were calculated using QRFCCA. We observed that one gene significantly influenced 6 traits; 2 genes, 5 traits; 4 genes, 4 traits; 6 genes, 3 traits; 17 genes, 2 traits and 20 genes, one trait. The remaining 17 genes did not reach the genome-wide significance with any trait. However, we observed that these genes still made genetic contributions to multiple traits. The significant association of the remaining 17 genes with the 46 traits was due to summation of the mild genetic effects on multiple traits of a single gene.

We also analyze the association of all common SNPs in one gene and one trait for all the 46 traits. The results were summarized in Table S25. We observed that 34 genes were significantly associated with creatinine, 29 genes with HsCRP, 23 genes with HOMA-IR, 23 genes with HOMA-B, 9 genes with glucose, 8 genes with insulin, 7 genes with GGT, 6 genes with triglycerides and 6 genes with VLDL. The distributions of the number of genes consisting of only common variants associated with traits were similar to that of rare variants although the number of genes associated with common variants was smaller than the number of genes associated with rare variants.

The literature confirmed many gene-trait associations which were identified in our study. Throughout this section, all genes included common variants only in the analysis. *REG1B*, which showed the most significant association with the 46 traits (p-value $\leq 1.65 \times 10^{-116}$), has been reported to be associated with glucose[71] (our analysis identified an association with p-value $\leq 1.31 \times 10^{-21}$), the production of insulin[72] (our analysis identified an association with insulin with p-value $\leq 1.81 \times 10^{-24}$ and HOMA-IR with p-value $\leq 5.02 \times 10^{-94}$), and triglyceride whose increment has deleterious effects on the function of islet beta cells[73] (our analysis showed an association with triglyceride with p-value $\leq 1.81 \times 10^{-24}$). *LEF1,* which was associated with the 46 traits (p-value $\leq 7.44 \times 10^{-83}$), has been related to insulin resistance[74] (the analysis demonstrated an association with HOMA-B with p-value $\leq 7.17 \times 10^{-52}$) and heart failure[75-77] (the analysis showed an association with HsCRP with p-value $\leq 1.22 \times 10^{-122}$ and Homocysteine with p-value $\leq 2.24 \times 10^{-14}$). *DYNC1H1*, which was associated with the 46 traits (p-value $\leq 3.46 \times 10^{-58}$), HOMA-IR (p-value $\leq 4.76 \times 10^{-108}$), insulin (p-value $\leq 2.83 \times 10^{-28}$), glucose (p-value $\leq 1.07 \times 10^{-10}$) and creatinine (p-value $\leq 2.57 \times 10^{-9}$), has been associated with hyperinsulinemia, hyperglycemia, the progress of glucose intolerance[78], and the blood creatinine level.[79] *DOCK7*, which presented associations with the 46 traits (p-value $\leq 4.42 \times 10^{-57}$), HsCRP (p-value $\leq 3.66 \times 10^{-53}$) and BMI (p-value $\leq 9.36 \times 10^{-7}$), has been associated with heart disease and ischemic stroke[80] and overweight and obesity.[81] Gene *GBF1*, which was associated with the 46 traits (p-value $\leq 6.30 \times 10^{-28}$), HOMA-B (p-value $\leq 2.91 \times 10^{-17}$), has been reported to be involved in insulin resistance and type 2 diabetes.[82] *METAP2*, which showed strong association with the 46 traits (p-value $\leq 2.62 \times 10^{-20}$), HOMA-B (p-value $\leq 3.14 \times 10^{-32}$), HOMA-IR (p-value $\leq 3.17 \times 10^{-17}$) and insulin (p-value $\leq 8.61 \times 10^{-10}$), has demonstrated associations with insulin resistance and insulin levels.[83] Gene *GRN*, which presented associations with the 46 traits (p-value $\leq 1.63 \times 10^{-16}$), HOMA-IR (p-value $\leq 2.28 \times 10^{-29}$) and insulin (p-value $\leq 7.89 \times 10^{-10}$), has been reported to associate with insulin resistance in type 2 diabetes patients[84] and the blood insulin levels.[85] Finally, gene *USP44*, which showed associations with the 46 traits (p-value $\leq 3.49 \times 10^{-14}$), HsCRP (p-value $\leq 1.81 \times 10^{-31}$) and HOMA-IR (p-value $\leq 7.75 \times 10^{-11}$), has been reported to associate with congenital heart disease[86], the increment of the HsCRP in congenital heart disease patients[87] and insulin resistance.[87]

**Discussion**

Investigating the pleiotropic effects of the genetic variants can provide important information to allow a deeper understanding of the complex genetic structures of health and disease. However, the identification of complete pleiotropic structures of high dimensional genotype-phenotypes poses great statistical and computational challenges. To meet these challenges, we have addressed several issues to overcome the critical barriers in advancing the development of novel statistical methods and computational algorithms for genetic pleiotropic analysis.

The first issue is to explore deep architectures of genotype-phenotype data in cross-phenotype association analysis. The traditional single trait and multiple trait analysis usually use genotype data in their raw form. These methods do not transform the raw data into a suitable internal representation in which association analysis can be used for distinguishing disease patterns from health patterns. However, the genetic variants include missense, nonsense, silent, insertions and deletions, frameshift, gain of function, loss of function, splicing site, transcription factor binding site, promotor, enhancer, histone modifications and regulatory motifs. DNA sequences and genetic variants are hierarchically organized in the genome. Exploring multiple levels of representation of the genetic variants and learning their internal patterns involved in the disease development can increase the power to detect the association of the genetic variants with phenotypes. To utilize the deep architectures of genetic variants across the genomic region, we proposed a new paradigm of association analysis that consists of three steps to combine multilevel data reduction and CCA. The first step is to apply FPCA to the original data for dimension reduction. The FPCA decomposes the genetic data into several functional components. Each functional component contains functional information of genetic variants across the genomic region, preserves the orders of genetic variants along the genomic and returns all possible pair-wise and high order linkage disequilibrium. If the phenotypes are function-valued physiological traits or RNA-seq data, the FPCA can also be applied to the phenotype data. The second step is to use quadratically regularized matrix factorization for further compressing the FPC scores into low rank representation and removing noisy data points. As a result, the FPCA and matrix factorization extracted useful genetic and phenotype information and deeply learned the internal genetic and phenotype representation. The third step is to apply CCA to the extracted FPC scores. Large scale simulations and real data analysis demonstrated that QRFCCA substantially outperformed all nine other statistics and FCCA outperformed the statistics without functional data analysis for data reduction.

The second issue is to develop a general framework for unifying association analysis, which provides a theoretic basis to evaluate various statistical methods for association analysis and design the guidance for developing novel statistics for testing the association of genetic variants with phenotypes. We used reproducing kernel Hilbert spaces (RKHS) as a general framework and the covariance operator as a general tool for unifying CCA, kernel CCA, functional CCA, dependence measure-based independence tests and other association analyses including GAMuT. We showed that the maximum heritability analysis and multivariate linear regression are equivalent to the classical CCA. Covariance is a key measure to assess linear association. Its extension to covariance operator provides a tool for quantifying the nonlinear association and derives kernel-based dependence measures and independence tests which form the basis for the GAMuT test. We also show that the KCCA is quite similar to the kernel independent test.

Finally, we considered the FCCA. To unify multivariate association tests and functional association tests, we used RKHS as a general framework for the formulation of the functional CCA. We showed that the dependence measured in the FPC score-based kernel analysis is asymptotically equal to the association measure of the FCCA. The FCCA also use the kernel-based dependence measure to develop association tests. Unlike the KCCA and GAMuT test where the kernels are selected by users, the FCCA uses the FPC scores as the feature space and derives the kernels from the data. This was why large-scale simulations and real data showed that in general, the FCCA outperformed the GAMuT and KCCA. Our FPC scores were generated via two steps. The first step used Fourier or wavelet expansions to derive eigenfunctions. The second step was to generate the FPC scores from the eigenfunctions. In the literature, some authors used one step FPCA to directly derive FPC scores from the Fourier or wavelet expansions. Our experiences showed that the number of functional principal components using two step FPCA was, in general, smaller than that directly derived from the Fourier or wavelet expansions. Therefore, two step FPCA had higher power than the one step FPCA.

The third issue is how to reveal pleiotropic structure of the genetic variants and quantify the degree of pleiotropy. Pleiotropy is a widely used word to indicate that a gene affects multiple traits. However, the nature and extent of pleiotropy is less precisely defined. Our approach to pleiotropy was based on the classical hypothesis-testing framework. We made formal statistical tests for the pleiotropy. The null model of the pleiotropy of a gene is the absence of any traits which the gene was associated with. The alternative model is the presence of at least one trait which the gene was associated with. Under the hypothesis-testing approach, the real data analysis showed that the global structure of pleiotropy consisted of three scenarios. The first scenario was the presence of a single trait which the gene was significantly associated with at the genome-wide significance level. The second scenario was the presence of multiple traits which the gene was significantly associated with at the genome-wide significance level. The third scenario was the presence of multiple traits which the gene was weakly associated with. In the first two scenarios, in addition to traits which the gene was significantly associated with, some traits that did not reach genome-wide level to have significant association with the gene, had weak associations with the gene at p-value $\leq 0.05$. In genetic pleiotropic studies, if we start with searching the genes significantly associated with a single trait at the genome-wide significance level, then many genes with pleiotropic effects will be missed. In the UK-10K real data analysis, single trait association analysis failed to identify 44 genes (27 genes with rare variants and 17 genes with common variants) with pleiotropic effects.

The fourth issue is the cross-phenotype association analysis with next-generation sequencing. The popular methods for cross-phenotype association analysis is to assess the influence of a single variant on multiple distinct phenotypes. These methods work very well for cross-phenotype association analysis of common variants, but are not suitable for testing the association of rare variants with multiple phenotypes. To illustrate the urgent need to develop gene-based statistical methods for cross-phenotype association analysis of rare variants, we searched the variants across the genome for significant associations with the multiple phenotypes. We found that 21,272 rare variants were significantly associated with the 46 traits at the genome-wide significance level after Bonferroni correction. It is highly unlikely that so many rare variants affected the 46 traits. To overcome this limitation, we developed the QRFCCA for gene-based cross-phenotype association analysis. The QRFCCA can be applied to both multivariate phenotypes, function-valued phenotypes and NGS genotype data. Since the genotype profiles of the common variants and rare variants have different patterns, to increase the power of the tests, we take the association tests of common variants and rare variants separately. We found that the significant genes with common variants only were not overlapped with the significant genes with rare variants only. In pleiotropic analysis, we should conduct cross-phenotype analysis for both common and rare variants and separately. The QRFCCA provides a powerful tool to accomplish this task.

To provide a guidance for cross phenotype association studies, we comprehensively evaluated the current existing statistics for cross-phenotype association by using large-scale simulations and real data analysis. We found that the proposed QRFCCA not only substantially outperformed all other widely used competing statistics, but also was very flexible. The QRFCCA can be used for association analysis of both common variants and rare variants, and any phenotypes including quantitative or qualitative, multivariate or function-valued phenotypes.

We performed cross-phenotype association analysis of a largest number of traits with NGS data up to the present time. We identified 79 genes including rare variants only which were significantly associated with the 46 traits and 67 genes including common variants only which were significantly associated with the 46 traits. These two sets of genes were not overlapped. Some of gene-phenotype association can be confirmed in the literature. We found that the largest number of the traits which a gene significantly affected at the genome-wide significance level was six and three in the cross-phenotype association analysis of common and rare variants, respectively. We also discovered that the largest number of traits which a gene affected with the P-value < 0.05 was 18 and 16 in the cross-phenotype association analysis of common and rare variants, respectively. In the single trait association analysis, we found that a large number of genes significantly affected creatinine (genes with rare variants: 345, genes with common variants: 34), HsCRP (genes with rare variants: 72, gene with common variants: 29) and HOMA-IR (genes with rare variants: 108, genes with common variants: 24).

The results presented in this paper are preliminary. The greatest lengths of the genes that were significantly associated with the 46 traits for rare and common variants in the real data analysis were 131Kb and 42Kb, respectively. The proposed methods may not have power to detect the association of the genes with lengths longer than these numbers. The number of basis functions for genotype profile expansion is an important factor for the power of the FPCA-based tests. We have not performed theoretical analysis to determine the appropriate number of basic functions for genotype profile expansions. We resort to ad hoc approaches to select the number of basis function in the expansions. The current pleiotropic analysis cannot identify the global causal structure of pleiotropy, which will decrease our power to unravel mechanisms underlying complex traits. To overcome this limitation, causal inference tools should be explored for cross-phenotype association analysis. The purpose of this paper is to stimulate further discussions regarding the great challenges we are facing in the pleiotropy analysis of high dimensional phenotypic and genomic data produced by modern sensors and next-generation sequencing.

**Description of Supplemental Data**

Supplemental Data include 1 note, 9 figures and 25 tables.

**Acknowledgments**

The project described was supported by Grant 1R01HL106034-01, and 1U01HG005728-01 from the National Institutes of Health. Thanks Dr. Peng Wei for providing the TwinsUK data. This study makes use of data generated by the UK10K Consortium, derived from samples from the TwinsUK datasets. A full list of the investigators who contributed to the generation of the data is available from www.uk10k.org.

**Web Resources**

A program for implementing the proposed methods can be downloaded from our website

http://www.sph.uth.tmc.edu/hgc/faculty/xiong/index.htm and http://www.bioconductor.org/

The UK10K data can be downloaded from http://www.uk10k.org.

**Figure Titles and Legends**

**Figure 1.** The power of curves as a function of sample sizes of 10 statistics for collectively testing the association of all rare variants in the gene with 10 traits with low correlations at the significance level $\alpha = 0.05$.

**Figure 2**. The power of curves as a function of sample sizes of 10 statistics for collectively testing the association of all rare variants in the gene with 10 traits with moderate correlations, respectively, at the significance level $\alpha = 0.05$.

**Figure 3.** The power of curves as a function of sample sizes of 10 statistics for collectively testing the association of all rare variants in the gene with 10 traits with high correlations, respectively, at the significance level $\alpha = 0.05$.

**Figure 4.** The power of curves as a function of sample sizes of 10 statistics for collectively testing the association of the gene including both rare and common variants with 10 traits with low correlations at the significance level $\alpha = 0.05$.

**Figure 5**. The power of curves as a function of sample sizes of 10 statistics for collectively testing the association of the gene including both rare and common variants with 10 traits with moderate correlations, respectively, at the significance level $\alpha = 0.05$.

**Figure 6.** The power of curves as a function of sample sizes of 10 statistics for collectively testing the association of the gene including both rare and common variants with 10 traits with high correlations, respectively, at the significance level $\alpha = 0.05$.

**Figure 7.** QQ plot of QRFCCA, FCCA, PCA and GAMuT with 95% confidence interval for rare variants. The negative logarithm of the observed ($y$ axis) and the expected ($x$ axis) P value is plotted for each gene (dot), and the red line indicates the null hypothesis of no true association.

**Figure 8.** Manhattan plot showing the genome-wide P values of association of the genes consisting of only rare variants with the 46 traits calculated using QRFCCA. The axis $x$ represented the chromosomal positions of 33,746 genes and axis $y$ showed their $-\log_{10} P$ values. The horizontal red line denotes the thresholds of $P = 1.48 \times 10^{-6}$ for genome-wide significance after Bonferroni correction.

**Figure 9.** Cross phenotype association heat map (46 traits/79 genes with rare variants only). The horizontal axis denotes the traits and the vertical axis denotes the genes. The color represented P-values. The smaller the P-value the deeper the red color.

**Figure 10.** QQ plot of QRFCCA, FCCA, PCA and GAMuT with 95% confidence interval for common variants. The negative logarithm of the observed ($y$ axis) and the expected ($x$ axis) P value is plotted for each gene (dot), and the red line indicates the null hypothesis of no true association.

**Figure 11.** Manhattan plot showing the genome-wide P values of association of the genes consisting of only common variants with the 46 traits calculated using QRFCCA. The axis $x$ represented the chromosomal positions of 33,746 genes and axis $y$ showed their $-\log_{10} P$ values. The horizontal red line denotes the thresholds of $P = 1.51 \times 10^{-6}$ for genome-wide significance after Bonferroni correction.

**Figure 12.** Cross phenotype association heat map (46 traits/67 genes with common variants only). The horizontal axis denotes the traits and the vertical axis denotes the genes. The color represented P-values. The smaller the P-value the deeper the red color.

**Tables**

Table 1. Type 1 error rates of 10 statistics for testing the association of rare variants in a gene with 15 highly correlated traits.

| Sample Size | Level | MANOVA | minP | PCA | CCA | USAT | GAMuT | FCCA | SCCA | QRFCCA | KCCA |
|---|---|---|---|---|---|---|---|---|---|---|---|
| 500 | 0.05 | 0.0494 | 0.0284 | 0.0495 | 0.0473 | 0.0399 | 0.0524 | 0.0477 | 0.0513 | 0.0492 | 0.0442 |
| | 0.01 | 0.0102 | 0.0058 | 0.0096 | 0.097 | 0.0074 | 0.0104 | 0.0097 | 0.0101 | 0.0099 | 0.0095 |
| | 0.001 | 0.0009 | 0.0007 | 0.0010 | 0.0010 | 0.0008 | 0.0010 | 0.0011 | 0.0010 | 0.0011 | 0.0010 |
| 1000 | 0.05 | 0.0512 | 0.0253 | 0.0501 | 0.0464 | 0.0595 | 0.0548 | 0.0514 | 0.0477 | 0.0497 | 0.0485 |
| | 0.01 | 0.0105 | 0.0053 | 0.0095 | 0.0107 | 0.0116 | 0.0106 | 0.0106 | 0.0099 | 0.0099 | 0.0099 |
| | 0.001 | 0.0010 | 0.0005 | 0.0010 | 0.0010 | 0.0011 | 0.0011 | 0.0010 | 0.0011 | 0.0010 | 0.0009 |
| 2000 | 0.05 | 0.0497 | 0.0271 | 0.0533 | 0.0496 | 0.0598 | 0.0526 | 0.0461 | 0.0486 | 0.0498 | 0.0417 |
| | 0.01 | 0.0106 | 0.0050 | 0.0099 | 0.0106 | 0.0123 | 0.0113 | 0.0104 | 0.0098 | 0.0096 | 0.0099 |
| | 0.001 | 0.0010 | 0.0005 | 0.0010 | 0.0009 | 0.0014 | 0.0011 | 0.0010 | 0.0010 | 0.0009 | 0.0009 |

Table 2. Type 1 error rates of 10 statistics for testing the association of both rare and common variants in a gene with 15 highly correlated traits.

| Sample Size | Level | MANOVA | minP | PCA | CCA | USAT | GAMuT | FCCA | SCCA | QRFCCA | KCCA |
|---|---|---|---|---|---|---|---|---|---|---|---|
| 500 | 0.05 | 0.0497 | 0.0276 | 0.0482 | 0.0473 | 0.0404 | 0.0492 | 0.0519 | 0.0516 | 0.0518 | 0.0490 |
| | 0.01 | 0.0094 | 0.0057 | 0.0097 | 0.0100 | 0.0076 | 0.0103 | 0.0105 | 0.0103 | 0.0095 | 0.0104 |
| | 0.001 | 0.0010 | 0.0006 | 0.0010 | 0.0010 | 0.0008 | 0.0010 | 0.0009 | 0.0009 | 0.0010 | 0.0009 |
| 1000 | 0.05 | 0.0484 | 0.0280 | 0.0495 | 0.0513 | 0.0579 | 0.0550 | 0.0467 | 0.0522 | 0.0503 | 0.0442 |
| | 0.01 | 0.0095 | 0.0055 | 0.0099 | 0.0097 | 0.0109 | 0.0105 | 0.0099 | 0.0099 | 0.0097 | 0.0096 |
| | 0.001 | 0.0010 | 0.0005 | 0.0010 | 0.0011 | 0.0012 | 0.0011 | 0.0010 | 0.0010 | 0.0010 | 0.0009 |
| 2000 | 0.05 | 0.0517 | 0.0266 | 0.0479 | 0.0534 | 0.0694 | 0.0556 | 0.0512 | 0.0530 | 0.0500 | 0.0481 |
| | 0.01 | 0.0099 | 0.0052 | 0.0096 | 0.0106 | 0.0125 | 0.0113 | 0.0096 | 0.0098 | 0.0102 | 0.0096 |
| | 0.001 | 0.0010 | 0.0005 | 0.0010 | 0.0010 | 0.0012 | 0.0011 | 0.0010 | 0.0010 | 0.0010 | 0.0010 |

Table 3. A list of P-values of 25 genes with rare variants only significantly associated with 46 traits using QRFCCA.

| Gene | Chr | Statistical Methods | | | | | | | | |
|---|---|---|---|---|---|---|---|---|---|---|
| | | QFCCA | FCCA | SCCA | PCA | KCCA | GAMuT | USAT | CCA | MANOVA |
| CTC-498M16.2 | 5 | 5.7E-22 | 2.9E-19 | 1.3E-03 | 2.6E-01 | 1.2E-06 | 1.9E-02 | 9.3E-01 | 4.7E-02 | 1.5E-02 |
| TRAJ22 | 14 | 2.2E-20 | 7.2E-18 | 1.0E-06 | 1.5E-12 | 8.6E-07 | 1.0E-05 | 8.6E-01 | 3.6E-01 | 2.6E-01 |
| AP000351.10 | 22 | 2.1E-18 | 3.9E-16 | 1.9E-03 | 7.4E-01 | 1.3E-06 | 2.5E-02 | 6.8E-01 | 2.8E-01 | 1.7E-01 |
| HAR1B | 20 | 7.8E-18 | 2.5E-15 | 1.0E-06 | 8.3E-02 | 3.1E-01 | 2.3E-01 | 8.4E-01 | 9.5E-01 | 9.9E-01 |
| IGHVII-20-1 | 14 | 7.5E-16 | 1.9E-15 | 1.2E-03 | 2.2E-01 | 4.5E-03 | 1.1E-01 | 9.5E-01 | 3.5E-01 | 1.1E-01 |
| RP11-4F5.2 | 15 | 9.9E-16 | 6.1E-13 | 2.1E-03 | 9.9E-01 | 3.6E-01 | 1.2E-04 | 9.7E-01 | 7.2E-01 | 4.7E-01 |
| RNVU1-17 | 1 | 3.9E-13 | 4.8E-13 | 1.2E-03 | 5.5E-02 | 3.0E-01 | 6.0E-06 | 8.2E-01 | 9.9E-01 | 7.4E-01 |
| PNOC | 8 | 1.6E-12 | 2.9E-12 | 1.0E-06 | 7.5E-01 | 3.1E-01 | 5.4E-05 | 8.8E-01 | 9.0E-02 | 3.9E-03 |
| COTL1P1 | 17 | 8.7E-12 | 1.4E-11 | 7.5E-04 | 6.2E-01 | 3.2E-06 | 5.4E-03 | 8.2E-01 | 1.5E-02 | 2.0E-03 |
| LINC00273 | 16 | 4.4E-11 | 1.8E-09 | 9.0E-04 | 6.5E-12 | 2.9E-03 | 1.7E-05 | 8.2E-01 | 7.1E-05 | 9.1E-06 |
| snoU13 | 12 | 5.0E-11 | 1.3E-10 | 6.9E-03 | 9.8E-01 | 1.1E-02 | 9.0E-03 | 8.7E-01 | 5.8E-01 | 2.2E-01 |
| ADAM19 | 5 | 6.1E-11 | 8.6E-09 | 1.0E-06 | 1.2E-01 | 6.9E-01 | 8.2E-01 | 8.8E-01 | 9.5E-01 | 9.7E-01 |
| CTD-2026G6.2 | 3 | 1.9E-10 | 9.3E-10 | 2.3E-03 | 6.4E-02 | 4.1E-01 | 3.2E-02 | 9.8E-01 | 9.3E-01 | 9.3E-01 |
| MIR409 | 14 | 2.8E-10 | 5.4E-10 | 4.1E-04 | 2.1E-03 | 5.1E-03 | 1.3E-01 | 1.0E+00 | 2.0E-02 | 8.9E-04 |
| RP1-276E15.1 | 11 | 3.1E-10 | 4.6E-10 | 5.2E-03 | 4.3E-01 | 8.5E-01 | 6.1E-03 | 8.7E-01 | 9.7E-01 | 9.5E-01 |
| HMGN1P6 | 2 | 4.5E-10 | 3.8E-08 | 1.0E-02 | 1.2E-02 | 4.7E-01 | 8.3E-06 | 8.2E-01 | 9.3E-01 | 9.8E-01 |
| HOXA7 | 7 | 2.2E-09 | 3.2E-09 | 1.0E-06 | 7.2E-01 | 9.9E-03 | 8.8E-02 | 9.4E-01 | 8.0E-01 | 4.2E-01 |
| RNA5SP99 | 2 | 2.4E-09 | 1.4E-09 | 2.1E-04 | 8.3E-01 | 6.5E-01 | 6.3E-04 | 9.4E-01 | 9.8E-01 | 6.9E-01 |
| AC021660.1 | 3 | 2.7E-09 | 2.5E-09 | 1.8E-04 | 1.3E-01 | 9.7E-01 | 7.8E-06 | 7.7E-01 | 4.4E-01 | 6.6E-02 |
| RP11-561N12.1 | 7 | 3.1E-09 | 9.1E-08 | 1.1E-04 | 2.1E-02 | 3.0E-01 | 5.5E-06 | 8.9E-01 | 8.4E-01 | 2.4E-01 |
| FBXL5 | 4 | 4.5E-09 | 9.9E-08 | 6.8E-04 | 9.2E-03 | 3.5E-01 | 6.6E-06 | 9.9E-01 | 9.9E-01 | 9.2E-01 |
| PPIAP23 | 13 | 5.4E-09 | 5.5E-09 | 4.3E-04 | 2.9E-05 | 4.0E-01 | 1.7E-05 | 1.0E+00 | 4.2E-01 | 5.3E-02 |
| HOXB2 | 17 | 7.0E-09 | 2.2E-09 | 1.0E-06 | 2.4E-02 | 6.3E-01 | 9.5E-06 | 1.0E+00 | 9.1E-01 | 8.4E-01 |
| RP11-170N16.1 | 4 | 7.1E-09 | 3.6E-07 | 6.2E-03 | 5.5E-05 | 9.4E-01 | 6.0E-04 | 1.0E+00 | 9.4E-01 | 9.2E-01 |
| AC008694.3 | 5 | 1.4E-08 | 5.0E-07 | 1.4E-03 | 1.1E-05 | 8.8E-01 | 2.4E-06 | 9.6E-01 | 9.9E-01 | 9.3E-01 |

Table 4. A list of P-value of 25 genes with common variants only significantly associated with 46 traits using QRFCCA.

| | | Statistical Methods | | | | | | | | |
|---|---|---|---|---|---|---|---|---|---|---|
| Gene | Chr | QRFCCA | FCCA | GAMuT | SCCA | USAT | MANOVA | CCA | PCA | KCCA |
| REG1B | 2 | 1.6E-116 | 7.4E-113 | 4.3E-01 | 1.3E-03 | 5.2E-01 | 1.0E+00 | 9.2E-01 | 3.0E-02 | 9.6E-01 |
| RP11-665C14.1 | 4 | 1.4E-93 | 8.2E-75 | 3.0E-01 | 4.6E-04 | 2.6E-01 | 9.8E-01 | 1.0E+00 | 3.9E-01 | 9.5E-01 |
| ZNF160 | 19 | 2.0E-91 | 6.7E-84 | 4.8E-02 | 5.6E-05 | 3.1E-01 | 9.0E-01 | 9.4E-01 | 1.3E-01 | 9.6E-01 |
| LEF1 | 4 | 7.4E-83 | 1.8E-82 | 2.8E-03 | 1.0E-06 | 7.0E-01 | 9.5E-01 | 9.7E-01 | 2.3E-03 | 9.9E-01 |
| DYNC1H1 | 14 | 3.5E-58 | 4.2E-48 | 5.4E-02 | 4.1E-03 | 6.6E-01 | 9.2E-01 | 9.4E-01 | 5.4E-01 | 9.4E-01 |
| DOCK7 | 1 | 4.4E-51 | 6.9E-50 | 1.7E-01 | 8.0E-06 | 9.4E-01 | 9.1E-01 | 9.7E-01 | 1.2E-01 | 9.5E-01 |
| SHC3 | 9 | 7.6E-42 | 2.2E-41 | 1.4E-02 | 4.2E-05 | 8.7E-02 | 9.7E-01 | 9.4E-01 | 1.3E-01 | 9.9E-01 |
| Y_RNA | 7 | 1.9E-36 | 1.3E-29 | 4.0E-03 | 3.5E-05 | 4.5E-01 | 1.0E+00 | 1.0E+00 | 7.9E-01 | 9.7E-01 |
| CTD-2122P11.1 | 5 | 1.6E-33 | 2.9E-31 | 7.4E-01 | 1.5E-05 | 2.3E-01 | 9.7E-01 | 9.8E-01 | 5.3E-01 | 9.9E-01 |
| GBF1 | 10 | 6.3E-28 | 2.3E-28 | 6.5E-02 | 4.5E-04 | 9.7E-01 | 9.7E-01 | 9.7E-01 | 8.4E-02 | 9.0E-01 |
| RP1-8B22.1 | 1 | 1.7E-27 | 4.1E-26 | 3.0E-02 | 7.3E-04 | 1.8E-01 | 1.0E+00 | 1.0E+00 | 5.0E-02 | 9.2E-01 |
| VPS13D | 1 | 2.6E-26 | 1.5E-27 | 7.9E-01 | 1.9E-03 | 8.6E-01 | 9.8E-01 | 1.0E+00 | 8.9E-01 | 9.4E-01 |
| RP11-68I3.2 | 17 | 3.2E-24 | 4.9E-24 | 2.2E-01 | 1.5E-05 | 1.0E+00 | 9.0E-01 | 9.6E-01 | 1.6E-02 | 9.9E-01 |
| SLC13A3 | 20 | 8.5E-24 | 2.0E-24 | 5.3E-01 | 9.0E-06 | 4.2E-01 | 9.9E-01 | 9.3E-01 | 7.9E-02 | 9.4E-01 |
| RP11-167N24.3 | 12 | 4.3E-23 | 1.0E-22 | 7.8E-01 | 1.1E-03 | 2.1E-01 | 9.2E-01 | 9.2E-01 | 6.3E-01 | 9.8E-01 |
| UBA6 | 4 | 5.3E-22 | 4.8E-22 | 2.0E-01 | 1.2E-03 | 4.9E-01 | 9.0E-01 | 1.0E+00 | 2.8E-01 | 9.2E-01 |
| GAN | 16 | 1.5E-21 | 1.0E-20 | 7.3E-02 | 3.0E-05 | 6.6E-01 | 9.7E-01 | 9.3E-01 | 2.9E-01 | 9.6E-01 |
| RP4-794H19.2 | 1 | 4.1E-21 | 1.3E-20 | 7.5E-01 | 1.6E-04 | 8.2E-02 | 9.3E-01 | 1.0E+00 | 9.1E-01 | 9.5E-01 |
| RP11-142I20.1 | 18 | 5.3E-21 | 8.5E-21 | 1.5E-01 | 4.3E-05 | 8.9E-01 | 9.3E-01 | 9.9E-01 | 3.4E-01 | 9.5E-01 |
| METAP2 | 12 | 2.6E-20 | 1.6E-20 | 6.8E-01 | 2.5E-05 | 5.2E-01 | 9.6E-01 | 9.4E-01 | 4.4E-01 | 9.4E-01 |
| SLCO1C1 | 12 | 5.9E-20 | 1.7E-19 | 7.5E-01 | 4.0E-06 | 3.8E-01 | 9.9E-01 | 9.1E-01 | 5.5E-02 | 9.1E-01 |
| AC105443.2 | 7 | 2.5E-17 | 3.4E-17 | 5.6E-02 | 1.2E-04 | 2.5E-01 | 9.6E-01 | 9.7E-01 | 2.6E-02 | 9.8E-01 |
| GRN | 17 | 1.6E-16 | 7.8E-16 | 5.0E-01 | 1.0E-03 | 1.2E-01 | 1.0E+00 | 1.0E+00 | 3.5E-02 | 9.6E-01 |
| INTS12 | 4 | 1.7E-16 | 1.2E-16 | 1.8E-02 | 1.0E-06 | 4.8E-01 | 9.7E-01 | 9.6E-01 | 3.3E-01 | 9.5E-01 |
| RP11-323I15.5 | 15 | 9.3E-16 | 2.0E-16 | 1.6E-01 | 5.0E-06 | 6.7E-01 | 9.7E-01 | 9.5E-01 | 3.5E-01 | 9.1E-01 |

**Supplemental Note**

**Supplemental Note A**

**Quantitative Genetics for Multiple Traits**

**Single Trait**

We first review quantitative genetics for a single trait. For simplicity, we only consider genetic additive effects. It is straightforward to extend analysis to include the genetic dominance effects. Consider the genetic model for a trait and a single locus:

$$Y = X\alpha + \varepsilon, \tag{N1}$$

where $Y$ denotes a trait, $X$ an additive genotype score, $\alpha$ the genetic additive effect and $\varepsilon$ error with mean zero and variance $\sigma_e^2$.

The genetic additive effect is estimated by

$$\hat{\alpha} = \frac{\text{cov}(x, y)}{\text{var}(x)}. \tag{N2}$$

The genetic additive variance $\sigma_A^2$ and narrow-sense heritability are

$$\sigma_A^2 = \text{var}(x\alpha) = \text{var}(x)\alpha^2 = 2pq\alpha^2 = \frac{[\text{cov}(x, y)]^2}{\text{var}(x)}, \tag{N3}$$

and

$$h^2 = \frac{\sigma_A^2}{\text{var}(Y)} = \frac{\text{var}(x\alpha)}{\text{var}(Y)} = \frac{[\text{cov}(x, y)]^2}{\text{var}(Y)\text{var}(x)} = corr^2(Y, X), \tag{N4}$$

respectively.

Equation (N4) shows that the narrow heritability can also be expressed as the proportion of the phenotype variation explained by the genetic variation or the squared correlation between the genotype and phenotype.

**Multiple Traits**

Quantitative genetics for a single trait can be easily extended to multiple traits. Again consider a genetic model for multiple traits and multiple loci:

$$Y = XB + E, \tag{N5}$$

where $Y = [Y_1,...,Y_k]$ represent a vector of $k$ phenotype variables, $X$ is a vector of $p$ genotype variables, $B = [b_1,...,b_k], b_j \in R^p$ the matrix of genetic additive effects and $\varepsilon = [\varepsilon_1,..., \varepsilon_k]$ is a vector of $k$ error variables.

The genetic effect matrix $B$ can be estimated by

$$\hat{B} = \Sigma_{xx}^{-1}\Sigma_{xy}. \tag{N6}$$

The covariance of the genetic additive effects is defined as

$$\begin{aligned}\Sigma_A &= \text{cov}(XB, XB) = B^T \Sigma_{xx} B \\ &= \Sigma_{yx} \Sigma_{xx}^{-1} \Sigma_{xy}.\end{aligned} \tag{N7}$$

The concept of heritability for a single trait is well developed. Recently, there has been an increasing attempt to extend heritability from a single trait to multiple traits. However, their focus is to consider the heritability of a linear combination of multiple traits.[50,59,60] Here, we consider a heritability matrix that is defined as

$$h_m^2 = \Sigma_{yy}^{-1/2} \Sigma_A \Sigma_{yy}^{-1/2}$$
$$= \Sigma_{yy}^{-1/2} \Sigma_{yx} \Sigma_{xx}^{-1} \Sigma_{xy} \Sigma_{yy}^{-1/2} .$$
(N8)

It is clear that the heritability matrix $h_m^2$ is equal to the matrix $R^2$ in equation (N8).

**Linear Combination of Multiple Traits**

We consider a linear combination of multiple phenotypes $Yb$ and a linear combination of genotypes at multiple loci $Xa$ to transform the association analysis of multiple traits to the association analysis of single trait. Define the linear genetic model for $Yb$ and $Xa$:

$$Yb = (Xa)\alpha + \varepsilon .$$
(N9)

The total genetic effect of the multiple genotypes on the multiple traits can be estimated by

$$\hat{\alpha} = \frac{\text{cov}(Xa, Yb)}{\text{var}(xa)} = \frac{a^T \Sigma_{xy} b}{a^T \Sigma_{xx} a} .$$
(N10)

Using equations (N3) and (N4), we obtain the genetic additive variance of $Xa$ and heritability of $yb$:

$$\sigma_{Al}^2 = \text{var}(Xa)\alpha^2 = \frac{(a^T \Sigma_{xy} b)^2}{a^T \Sigma_{xx} a} ,$$
(N11)

and

$$h_l^2 = \frac{\text{var}(Xa)\alpha^2}{\text{var}(Yb)} = \frac{(a^T \Sigma_{xy} b)^2}{b^T \Sigma_{yy} b a^T \Sigma_{xx} a} ,$$
(N12)

respectively.

It is clear that the squared multiple correlation coefficient is given by

$$R^2 = \frac{[\text{cov}(Yb,(Xa)\alpha]^2}{\text{var}(Yb)\text{var}((Xa)\alpha)} = \frac{[\text{cov}(Yb,Xa)]^2 \alpha^2}{\text{var}(Yb)\text{var}(Xa)\alpha^2} = \frac{(a^T \Sigma_{xy} b)^2}{b^T \Sigma_{yy} b a^T \Sigma_{xx} a} = h_l^2.$$

Next we seek the optimal combinations of the genotypes at multiple loci and the multiple traits to maximize the genetic additive effect, genetic additive variance and heritability. We first find the maximum genetic additive effect.

Using equation (N10) and the Lagrangian multiplier method, we can solve the following optimization problem to obtain the maximum genetic additive effect:[50]

$$L(a,b,\lambda) = a^T \Sigma_{xy} b + \frac{\lambda}{2}(1 - a^T \Sigma_{xx} a),  \qquad (N13)$$

where $\lambda$ is a multiplier.

Setting $\dfrac{\partial L(a,b,\lambda)}{\partial a} = \Sigma_{xy} b - \lambda \Sigma_{xx} a = 0$ gives

$\Sigma_{xy} b = \lambda \Sigma_{xx} a$ or

$$\Sigma_{xx}^{-1} \Sigma_{xy} b = \lambda a. \qquad (N14)$$

Suppose that the SVD of the matrix $\Sigma_{xx}^{-1} \Sigma_{xy}$ is given by

$$\Sigma_{xx}^{-1} \Sigma_{xy} = U_e \Lambda_e V_e^T. \qquad (N15)$$

Then, it follows from equation (N15) that $a$, $b$ and the optimal genetic effect are the left, right singular vectors and singular value of $\Sigma_{xx}^{-1} \Sigma_{xy}$, respectively. Similarly, we can show that the genetic additive variance $\sigma_{Al}^2$ is the square of the singular value of $\Sigma_{xx}^{-1} \Sigma_{xy}$. The maximum heritability can be obtained by setting the Lagrange function:

$$L(a,b,\lambda,\mu) = (a^T\Sigma_{xy}b)^2 + \lambda(1 - a^T\Sigma_{xx}a) + \mu(1 - b^T\Sigma_{yy}b) \text{ and}$$

$$\frac{\partial L}{\partial a} = 2(a^T\Sigma_{xy}b)\Sigma_{xy}b - 2\lambda\Sigma_{xx}a = 0$$
$$\frac{\partial L}{\partial b} = 2(a^T\Sigma_{xy}b)\Sigma_{yx}a - 3\mu\Sigma_{yy}b = 0.$$

(N16)

Let

$K = \Sigma_{xx}^{-1/2}\Sigma_{xy}\Sigma_{yy}^{-1/2}$ and SVD of $K$ be

$$\Sigma_{xx}^{-1/2}\Sigma_{xy}\Sigma_{yy}^{-1/2} = U_K \Lambda_K V_K^T.$$

(N17)

Solving equation (N16) we obtain

$$a = \Sigma_{xx}^{-1/2} u_k$$
$$b = \Sigma_{yy}^{-1/2} v_k$$
$$h_l^2 = \lambda = \tau_K^2,$$

(N18)

where $u_k$, $v_k$ and $\tau_K$ are the left, right singular vectors and singular value of the matrix $K$, respectively. Substituting equation (N18) into equations (N10) and (N11) gives

$$\hat{\alpha} = u_K^T K v_K$$
$$\sigma_{Al}^2 = \hat{\alpha}^2.$$

(N19)

Note that $\tau_K^2$ is the eigenvalue of the matrix

$$K^T K = \Sigma_{yy}^{-1/2}\Sigma_{yx}\Sigma_{xx}^{-1}\Sigma_{xy}\Sigma_{yy}^{-1/2}.$$

(N20)

It follows from equation (N20) and (18) that

$$K^T K = R^2.$$

This shows that the maximum heritability analysis is equivalent to CCA.

**Supplemental Note B**

**RKHS framework for Functional CCA, Cross-covariance operator, Dependence Measure and Independence Test**

Many multivariate and functional statistical methods such as regression, CCA, kernel regression, kernel CCA, functional regression and functional CCA, dependence measure can be used to test the association of genetic variants with the phenotypes. In the past decades, reproducing kernel Hilbert spaces (RKHS) have emerged as a general framework for various statistical and machine learning methods. Here, we propose to use RKHS as a unified framework for association tests to reveal the relationships among various multivariate and functional association tests.

We begin with briefly introducing RKHS.[1,2] Let $\mathcal{H}$ be a Hilbert space of functions on a non-empty set $\chi$ and denote the inner product in $\mathcal{H}$ by $<.,.>_{\mathcal{H}}$. A bivariate function $K$ on $\chi \times \chi$ is called a reproducing kernel for $\mathcal{H}$ if $K$ satisfies

(1) For every $t \in \chi$, $K(.,t) \in \mathcal{H}$ and

(2) For every $t \in \chi$ and $f \in \mathcal{H}$, we have $f(t) = <f, k(.,t)>_{\mathcal{H}}$.

We call $\mathcal{H}$ a RKHS with reproducing kernel $K$.

Two approaches can be used to construct a RKHS. The first approach is by completion of the space of all functions of the form $\sum_{i=1}^{n} \alpha_i K(.,s_i)$, $\alpha_i \in R, s_i \in \chi, n = 1,2,...,$ with the inner product defined as

$$<\sum_{i=1}^{n} \alpha_i K(.,s_i), \sum_{j=1}^{l} \beta_j K(.,t_j)>_{\mathcal{H}(K)} = \sum_{i=1}^{n}\sum_{j=1}^{l} \alpha_i \beta_j K(s_i,t_j). \tag{N21}$$

The constructed RKHS with reproducing kernel $K$ is denoted by $\mathcal{H}(\mathcal{K})$.

The second approach is via completion of the set of all linear combinations of random variables. Consider a stochastic process $\{X(t): t \in \chi\}$ with mean zero where $X(t)$ denotes a map from a probability space into $R$ and $E[X(t)] = 0$ for all $t \in \chi$. We define $L_X^2$ as the completion of the set of all linear combinations of the random variables of the form:

$$\sum_{i=1}^{n} \alpha_i X(t_i), \alpha_i \in R, t_i \in \chi, n = 1, 2, \ldots, \tag{N22}$$

with the inner product $< X(t), X(s) >_{L_X^2} = E[X(t)X(s)]$.

Let $K(s,t) = E[X(s)X(t)]$ be a covariance function which generates a RKHS. Two approaches can be connected by the inner product:

$$\begin{aligned} \|\sum_i \alpha_i X(t_i)\|_{L_X^2}^2 &= E[(\sum_i \alpha_i X(t_i))^2] \\ &= \sum_i \sum_j \alpha_i \alpha_j E[X(t_i)X(t_j)] \\ &= \sum_i \sum_j \alpha_i \alpha_j K(t_i, t_j) \\ &= \sum_i \sum_j \alpha_i \alpha_j < K(t_i,.), K(t_j,.) >_{\mathcal{K}(\mathcal{K})} \\ &= \|\sum_i \alpha_i K(.,t_i)\|_{\mathcal{H}(\mathcal{K})}^2 . \end{aligned} \tag{N23}$$

A powerful analytical tool for CCA and independence tests is the cross-covariance operator that is an extension of the covariance matrix to infinite dimensional space.[3] Recall that the covariance matrix is defined as

$$\tilde{\Sigma}_{XY} = E[XY^T], \tag{N24}$$

where $X$ and $Y$ are vectors of random variables with $E[X]=0$ and $E[Y]=0$. Equation (N24) can be extended to feature space. Let $\phi(X)$ and $\psi(Y)$ be feature maps. In the feature space, equation (N24) can be written as

$$\tilde{\Sigma}_{XY} = E[\phi(X)\psi^T(Y)]. \tag{N25}$$

Let $f$ and $g$ be vectors in the feature space. Recall that by the reproducing property, we have

$$f(X) = <f(.), K(.,X)> \text{ and } g(Y) = <g(.), K(.,Y)>. \tag{N26}$$

Define kernels $K(.,X) = \phi(X)$ and $K(.,Y) = \psi(Y)$. Viewing the covariance matrix $\tilde{\Sigma}_{XY}$ as an operator and applying it to the vector $g$, we obtain

$$\begin{aligned}\tilde{\Sigma}_{XY} g &= E[\phi(X)\psi^T(Y)g(.)] = E[\phi(X) < K(.,Y), g(.) >] \\ &= E[\phi(X)g(Y)] \\ &= E[\phi(X)g(Y)].\end{aligned} \tag{N27}$$

Equation (N27) indicates that $\tilde{\Sigma}_{XY} g$ maps $g$ to a vector in the feature space spanned by $\phi(X)$.

Let $f$ be a vector in the feature space. Then, its inner product with $\tilde{\Sigma}_{XY} g$ is given by

$$\begin{aligned}f^T\tilde{\Sigma}_{XY} g &= E[f^T\phi(X)g(Y)] \\ &= E[<f(.), K(.,X)> g(Y)]. \\ &= E[f(X)(g(Y)].\end{aligned} \tag{N28}$$

In terms of kernels, equation (N28) can be written as

$$f^T\tilde{\Sigma}_{XY} g = E[<f(.), K(.,X)><K(Y,.), g(.)>]. \tag{N29}$$

We assume that

$$f = \sum_{i=1}^{m} \alpha_i \phi(X_i) \text{ and } g = \sum_{j=1}^{m} \beta_j \psi(Y_j).  \qquad (N30)$$

From equations (N27) and (N28) we can obtain the sampling formula for $f^T \tilde{\Sigma}_{XY} g$:

$$f^T \tilde{\Sigma}_{XY} g = \frac{1}{m} \sum_{i=1}^{m} < f(.), K(., X_i) >< K(Y_i,.), g(.) >. \qquad (N31)$$

Substituting equation (N30) into equation (N31) gives

$$\begin{aligned}f^T \tilde{\Sigma}_{XY} g &= \frac{1}{m} \sum_{i=1}^{m} < \sum_{j=1}^{m} \alpha_j K(X_j,.), K(., X_i) >< K(Y_i,.), \sum_{l=1}^{m} \beta_l K(Y_l,.) > \\ &= \frac{1}{m} \sum_{j=1}^{m} \sum_{l=1}^{m} \alpha_j \beta_l \sum_{i=1}^{m} K(X_j, X_i) K(Y_i, Y_l) \qquad (N32)\\ &= \frac{1}{m} \alpha^T K_x K_y \beta,\end{aligned}$$

where

$$K_x = \begin{bmatrix} K(X_1, X_1) & \cdots & K(X_1, X_m) \\ \vdots & \vdots & \vdots \\ K(X_m, X_1) & \cdots & K(X_m, X_m) \end{bmatrix} \text{ and } K_y = \begin{bmatrix} K(Y_1, Y_1) & \cdots & K(Y_1, Y_m) \\ \vdots & \vdots & \vdots \\ K(Y_m, Y_1) & \cdots & K(Y_m, Y_m) \end{bmatrix}.$$

Let $G = I_m - \frac{1}{m} \mathbf{1}_m$, where $\mathbf{1}_m$ is a $m \times m$ matrix of ones. The centered covariance is

$$\Sigma_{XY} = \tilde{\Sigma}_{XY} - \mu_X \mu_Y^T,$$

where $\mu_X = E[\phi(X)]$ and $\mu_Y = E[\psi(Y)]$.

Using the similar arguments, we can show

$$f^T \Sigma_{XY} g = \frac{1}{m} \alpha^T \tilde{K}_x \tilde{K}_y \beta, \qquad (N33)$$

where $\tilde{K}_x = G K_x G$ and $\tilde{K}_y = G K_y G$.

The covariance operator is a useful tool for assessing dependence between variables and hence form a foundation for association analysis. A dependence measure can be derived from solving the following optimization problem:

$$\max_{\alpha,\beta} \quad \frac{1}{m}\alpha^T \tilde{K}_x \tilde{K}_y \beta$$
$$\text{s.t.} \quad \alpha^T \tilde{K}_x \alpha = 1 \tag{N34}$$
$$\beta^T \tilde{K}_y \beta = 1.$$

Let $u = \tilde{K}_x^{1/2}\alpha$ and $v = \tilde{K}_y^{1/2}$. Then, the optimization problem (N34) can be transformed to

$$\max_{u,v} \quad \frac{1}{m} u^T \tilde{K}_{xx}^{1/2} \tilde{K}_{yy}^{1/2} v$$
$$\text{s.t.} \quad u^T u = 1 \tag{N35}$$
$$v^T v = 1.$$

Using the Lagrange multiplier approach to solve the optimization problem, we can obtain the eigenequation:

$$\frac{1}{m}\tilde{K}_{xx}^{1/2}\tilde{K}_{yy}^{1/2} v = \lambda u, \tag{N36}$$

$$\frac{1}{m}\tilde{K}_{yy}^{1/2}\tilde{K}_{xx}^{1/2} u = \lambda v. \tag{N37}$$

Substituting equation (N37) into equation (N36) gives eigenequation:

$$\frac{1}{m^2}\tilde{K}_{xx}^{1/2}\hat{K}_{yy}\tilde{K}_{xx}^{1/2} u = \lambda^2 u. \tag{N37}$$

Assume that the single value decomposition of $\tilde{K}_{xx}^{1/2}\tilde{K}_{yy}^{1/2}$ is

$$\tilde{K}_{xx}^{1/2}\tilde{K}_{yy}^{1/2} = U\Lambda V^T, \tag{N38}$$

where $\Lambda = \text{diag}(\rho_1, \rho_2, ..., \rho_m)$ with $\rho_1 \geq \rho_2 \geq ... \geq \rho_m$.

After some algebra, we obtain

$$\frac{1}{m^2}\sum_{i=1}^{n}\rho_i^2 = \frac{1}{m^2}\text{Trace}(U^T \tilde{K}_{xx}^{1/2} \tilde{K}_{yy} \tilde{K}_{xx}^{1/2} U)$$
$$= \frac{1}{m^2}\text{Trace}(\tilde{K}_{xx}\tilde{K}_{yy}).$$
(N39)

It is clear that $\frac{1}{m^2}\text{Trace}(\tilde{K}_{xx}\tilde{K}_{yy})$ can be used as a measure of dependence. The dependence measure $\frac{1}{m^2}\text{Trace}(\tilde{K}_{xx}\tilde{K}_{yy})$ is used to test for dependence between two sets of random variables[4] and to test for the association of genetic variants with multiple phenotypes.[5]

Consider Kernel CCA (KCCA).[9] Let $\Phi(X) = [\phi(X_1),...,\phi(X_m)]^T$ and $\Psi(Y) = [\psi(Y_1),...,\psi(Y_m)]^T$. The representation of linear combinations of the sampled features in the feature space can be defined as

$$U = \Phi^T(X)G\alpha \text{ and } V = \Psi^T(Y)G\beta.$$

The canonical variates are linear combinations of features and hence can be represented by

$$a = G\Phi(X)U = G\Phi(X)\Phi^T(X)G\alpha = \tilde{K}_x \alpha \text{ and } b = G\Psi(Y)V = G\Psi(Y)\Psi^T(Y)G\beta = \tilde{K}_y \beta,$$

where $\tilde{K}_x = G\Phi(X)\Phi^T(X)G$ and $\tilde{K}_y = G\Psi(Y)\Psi^T(Y)G$. The covariance between the canonical variates $a$ and $b$ is $\text{cov}(a,b) = \alpha^T \tilde{K}_x \tilde{K}_y \beta$. Similarly, we can obtain the variance of $a$ and the variance of $b$, respectively, $\text{var}(a) = \alpha^T \tilde{K}_x \tilde{K}_x \alpha$ and $\text{var}(b) = \beta^T \tilde{K}_y \tilde{K}_y \beta$. The KCCA seeks canonical vectors in terms of $\alpha$ and $\beta$ to optimize

$$\max_{\alpha,\beta} \quad \frac{1}{m}\alpha^T \tilde{K}_x \tilde{K}_y \beta$$
$$\text{s.t.} \quad \alpha^T \tilde{K}_x \tilde{K}_x \alpha = 1$$
$$\beta^T \tilde{K}_y \tilde{K}_y \beta = 1.$$
(N40)

If the constraints in (N40) are replaced by $Var(U) = \alpha^T \tilde{K}_x \alpha = 1$ and $var(V) = \beta^T \tilde{K}_y \beta = 1$, the optimization problem (N40) can be reduced to

$$\max_{\alpha,\beta} \quad \frac{1}{m} \alpha^T \tilde{K}_x \tilde{K}_y \beta$$
$$\text{s.t.} \quad \alpha^T \tilde{K}_x \alpha = 1 \quad \quad \quad \quad \quad \quad \quad \quad \text{(N41)}$$
$$\beta^T \tilde{K}_y \beta = 1,$$

which is exactly the same as the formulation (N34). Similar to equation (22), the association measure in the KCCA is equal to

$$\frac{1}{m^2} \text{Tr}(\Lambda^2) = \frac{1}{m^2} \text{Trace}(\hat{K}_{xx} \hat{K}_{yy}). \quad \quad \text{(N42)}$$

To unify multivariate association tests and functional association tests, we use RKHS as a general framework for formulation of the functional CCA.[1] Consider two index sets $E_1$ and $E_2$, and two stochastic processes: $\{X(t), t \in E_1\}$ and $\{Y(s), s \in E_2\}$ with mean zero $E[X(t)] = E(Y(s)) = 0$ for all $t, s$, auto covariance functions

$R_X(s_1, s_2) = cov(X(s_1), X(s_2))$, $R_Y(t_1, t_2) = cov(Y(t_1), Y(t_2))$ and cross covariance functions

$R_{XY}(s,t) = cov(X(s), Y(t))$.

Let $L_X^2$ and $L_Y^2$ be the Hilbert spaces spanned by the $X$ and $Y$ processes defined as the completion of the set of all linear combinations of the random variables:

$$\{a : a = \sum_{i=1}^{n} \alpha_i X(t_i), t_i \in E_1, \alpha_i \in R, n \in Z^+\} \text{ and}$$

$\{b : b = \sum_{i=1}^{n} \beta_i X(s_i), s_i \in E_2, \beta_i \in R, n \in Z^+\}$ under the inner product $<a_1, a_2>_{L_X^2} = E[a_1 a_2]$ and

$<b_1, b_2>_{L_Y^2} = E[b_1 b_2]$, respectively. The covariance kernel can be used to define an integral operator:

$$(R_X \beta)(t) = \int_T R_X(t,s) \beta(s) ds .\tag{N43}$$

By Mercer's theorem[6], the covariance kernel function can be expanded in terms of orthonormal functions:

$$R_X(s,t) = \sum_{i=1}^{\infty} \lambda_i \phi_i(s) \phi_i(t) ,\tag{N44}$$

$$R_Y(s,t) = \sum_{j=1}^{\infty} \mu_j \theta_j(s) \theta_j(t) ,\tag{N45}$$

where

$(R_X \phi_i)(t) = \lambda_i \phi_i(t)$ and

$(R_Y \theta_j)(t) = \mu_j \theta_j(t)$.

Using the Karhunen-Loeve representation, the stochastic processes $X(t)$ and $Y(t)$ can be expressed as[1]

$$X(t) = \sum_{i=1}^{\infty} \xi_i \phi_i(t) \text{ and } Y(t) = \sum_{j=1}^{\infty} \eta_j \theta_j(t) ,\tag{N46}$$

where $\xi_i$ and $\eta_j$ are uncorrelated variables with zero means, and variances and covariance

$$\text{var}(\xi_i) = \lambda_i, \text{ var}(\eta_j) = \mu_j, \text{ cov}(\xi_i, \eta_j) = \gamma_{ij} ,\tag{N47}$$

which implies

$$R_{XY}(s,t) = \sum_{i=1}^{\infty}\sum_{j=1}^{\infty} \gamma_{ij}\phi_i(s)\theta_j(t). \tag{N48}$$

The RKHS $\mathcal{H}(R_X)$ generated by the covariance kernel $R_X$ is

$$\mathcal{H}(R_X) = \{\alpha : \alpha = \sum_{j=1}^{\infty}\lambda_j\alpha_j\phi_j(t), \sum_{j=1}^{\infty}\lambda_j\alpha_j^2 < \infty\}. \tag{N49}$$

Similarly, we can define

$$\mathcal{H}(R_Y) = \{\beta : \beta = \sum_{j=1}^{\infty}\mu_j\beta_j\theta_j(t), \sum_{j=1}^{\infty}\lambda_j\beta_j^2 < \infty\}. \tag{N50}$$

The congruence mapping from $\mathcal{H}(R_X)$ to $L_X^2$ is then given by[7]

$$\Psi_X(\alpha) = \sum_{i=1}^{\infty}\lambda_i\alpha_i <X,\phi_i>_{L^2(T)}. \tag{N51}$$

Similarly, we define

$$\Psi_Y(\alpha) = \sum_{j=1}^{\infty}\mu_j\beta_j <Y,\theta_j>_{L^2(T)}. \tag{N52}$$

Using equations (N51) and (N52), we obtain

$$\mathrm{cov}(\Psi_X(\alpha),\Psi_Y(\beta)) = \sum_{i=1}^{\infty}\sum_{j=1}^{\infty}\lambda_i\mu_j\alpha_i\beta_j\,\mathrm{cov}(<X,\phi_i>_{L^2(T)},<Y,\theta_j>_{L^2(T)}). \tag{N53}$$

By definition of the inner product in the $L^2(T)$ space, we have

$$<X,\phi_i>_{L^2(T)} = \int_T X(t)\phi_i(t)dt \text{ and } <Y,\theta_j>_{L^2(T)} = \int_T Y(t)\theta_j(t)dt. \tag{N54}$$

Then, using stochastic integral theory,[8] we can connect the covariance between these two inner products with the covariance operator:

$$\mathrm{cov}(<X,\phi_i>_{L^2(T)},<Y,\theta_j>_{L^2(T)}) = \int_T\int_T \phi_i(s)R_{XY}(s,t)\theta_j(t)dsdt. \tag{N55}$$

Substituting equation (N55) into equation (N53) gives

$$\begin{aligned}\operatorname{cov}(\Psi_X(\alpha),\Psi_Y(\beta)) &= \sum_{i=1}^{\infty}\sum_{j=1}^{\infty}\lambda_i\mu_j\alpha_i\beta_j\int_T\int_T \phi_i(s)R_{XY}(s,t)\theta_j(t)dsdt \\ &= \int_T\int_T (\sum_{i=1}^{\infty}\lambda_i\alpha_i\phi_i(s))R_{XY}(s,t)(\sum_{j=1}^{\infty}\mu_j\beta_j\theta_j)dsdt \\ &= \int_T\int_T \alpha(s)R_{XY}(s,t)\beta(t)dsdt.\end{aligned} \qquad (N56)$$

Substituting equation (N46) into equation (N56) yields

$$\operatorname{cov}(\Psi_X(\alpha),\Psi_Y(\beta)) =<<(R_{XY}\alpha)(t),\beta(t)>_{L^2(T)}, \qquad (N57)$$

where $R_{XY}$ with the property $(R_{XY}\alpha)(t)=\int_T R_{XY}(s,t)\alpha(s)ds=<R_{XY}(.,t),\alpha(.)>_{L^2(T)}$ is a cross covariance operator. Equation (N57) is an extension of equation (N28) to functional space.

Now we formally define CCA for the stochastic process. The canonical correlation can be defined in terms of both $L_X^2$ and $L_Y^2$, and $\mathcal{H}(R_X)$ and $\mathcal{H}(R_Y)$:

$$\begin{aligned}&\max_{\substack{\zeta\in L_X^2,\upsilon\in L_Y^2 \\ \operatorname{var}(\zeta)=1,\operatorname{var}(\upsilon)=1}} \operatorname{cov}^2(\zeta,\upsilon) \\ &= \max_{\substack{\alpha\in\mathcal{H}(R_X),\beta\in\mathcal{H}(R_Y) \\ \|\alpha\|^2_{\mathcal{H}(R_X)}=1,\|\beta\|^2_{\mathcal{H}(R_Y)}=1}} \operatorname{cov}^2(\Psi_X(\alpha),\Psi_Y(\beta))\end{aligned}. \qquad (N58)$$

Now we consider the direct extension of the CCA from multivariate to functional space. Suppose that expansions of the two functions $\alpha(s)$ and $\beta(t)$ in terms of orthonormal functions $\{\phi_i(s), i=1,2,...\}$ and $\{\theta_j(t), j=1,2,...\}$ are given by

$$\alpha(s) = \sum_{i=1}^{\infty}\alpha_i\phi_i(s) \text{ and } \beta(t) = \sum_{j=1}^{\infty}\beta_j\theta_j(t). \qquad (N59)$$

Define the inner product of functions $\alpha(s)$ with stochastic process $X(s)$ as

$$<\alpha(s),X(s)>_{L^2(T)} = \int_T \alpha(s)X(s)ds. \qquad (N60)$$

Substituting equation (N43) into equation (N60), we obtain

$$<\alpha(s), X(s)>_{L^2(T)} = \sum_{i=1}^{\infty} \xi_i \int_T \alpha(s)\phi_i(s)ds$$
$$= \sum_{i=1}^{\infty} \xi_i \alpha_i,$$

(N61)

where $\alpha_i = \int_T \alpha(s)\phi_i(s)ds$.

Similarly, we have

$$<\beta(t), Y(t)>_{L^2(T)} = \sum_{j=1}^{\infty} \eta_j \beta_j, \quad \beta_j = \int_T \beta(t)\theta_j(t)dt.$$

(N62)

Similar to multivariate CCA where we consider the correlation between linear combinations of variables in two sets, the functional CCA consider extensions of linear combinations of the random variables to the functional space. Using equations (N48), (N61) and (N62) we calculate the covariance:

$$\text{cov}(\int_T \alpha(s)X(s)ds, \int_T \beta(t)Y(t)dt) = \int_T \alpha(s) R_{XY}(s,t)\beta(t)dt\gamma$$
$$= \sum_{i=1}^{\infty}\sum_{j=1}^{\infty} \alpha_i \beta_j \gamma_{ij}$$

(N63)

Similarly, we have

$$\text{var}(\int_T \alpha(s)X(s)ds) = \sum_{i=1}^{\infty} \lambda_i \alpha_i^2$$
$$\text{var}(\int_T \beta(t)Y(t)dt = \sum_{j=1}^{\infty} \mu_j \beta_j^2.$$

(N64)

Therefore, the FCCA can be defined as

$$\max_{\alpha,\beta} \quad \alpha^T \Lambda_{XY} \beta$$
$$\text{s.t.} \quad \alpha^T \Lambda_{XX} \alpha = 1, \beta^T \Lambda_{YY} \beta = 1,$$

(N65)

where

$$\Lambda_{XY} = \begin{bmatrix} \gamma_{11} & \cdots & \gamma_{1q} \\ \vdots & \vdots & \vdots \\ \gamma_{p1} & \cdots & \gamma_{pq} \end{bmatrix}, \Lambda_{XX} = \begin{bmatrix} \lambda_1 & \cdots & 0 \\ \vdots & \vdots & \vdots \\ 0 & \cdots & \lambda_p \end{bmatrix}, \Lambda_{YY} = \begin{bmatrix} \mu_1 & \cdots & 0 \\ \vdots & \vdots & \vdots \\ 0 & \cdots & \mu_q \end{bmatrix}, \alpha = \begin{bmatrix} \alpha_1 \\ \vdots \\ \alpha_p \end{bmatrix}, \beta = \begin{bmatrix} \beta_1 \\ \vdots \\ \beta_q \end{bmatrix}.$$

Comparing equation (N63) with equation (N56) gives

$$\mathrm{cov}(\int_T \alpha(s)X(s)ds, \int_T \beta(t)Y(t)dt) = \mathrm{cov}(\Psi_X(\alpha), \Psi_Y(\beta)).$$

This shows that the formulation of FCCA in equation (N65) is equivalent to the formulation of the FCCA in the $L_X^2, L_Y^2$ and $\mathcal{H}(R_X), \mathcal{H}(R_Y)$ as expressed in equation (N58). Recall that

$$R^2 = \Lambda_{YY}^{-1/2} \Lambda_{YX} \Lambda_{XX}^{-1} \Lambda_{XY} \Lambda_{YY}^{-1}.$$

After some algebra, we obtain the association measure $r$ in equation (22):

$$r = \mathrm{Tr}(R^2) = \sum_{i=1}^{p} \sum_{j=1}^{q} \frac{\gamma_{ij}^2}{\lambda_i \mu_j}. \tag{N66}$$

If constraints $\alpha^T \Lambda_{XX} \alpha = 1, \beta^T \Lambda_{YY} \beta = 1$ in equation (N65) is replaced by $\alpha^T \alpha = 1$ and $\beta^T \beta = 1$, i.e., the optimization problem (N65) is reduced to

$$\begin{aligned} \max_{\alpha, \beta} \quad & \alpha^T \Lambda_{XY} \beta \\ \mathrm{s.t.} \quad & \alpha^T \alpha = 1, \beta^T \beta = 1, \end{aligned} \tag{N67}$$

then equation (N66) is reduced to

$$r = \mathrm{Tr}(R^2) = \sum_{i=1}^{p} \sum_{j=1}^{q} \gamma_{ij}^2. \tag{N68}$$

Assume that the expansion of $X(t)$ and $Y(t)$ in equation (N46) are truncated by $p$ and $q$ terms, respectively. Define the feature maps:

$\Phi(X_i) = [\xi_{i1}, \xi_{i2}, ..., \xi_{ip}]^T = \xi_i$ and $\Phi(Y_i) = [\eta_{i1}, \eta_{i2}, ..., \eta_{iq}]^T = \eta_i$, where $\xi_{ik}$ and $\eta_{il}$ are the FPC score of $X_i(t)$ and $Y_i(t)$, respectively.

Let

$$\Phi(X) = \begin{bmatrix} \Phi^T(X_1) \\ \vdots \\ \Phi^T(X_m) \end{bmatrix} = \xi = \begin{bmatrix} \xi_{11} & \cdots & \xi_{1p} \\ \vdots & \vdots & \vdots \\ \xi_{m1} & \cdots & \xi_{mp} \end{bmatrix} \text{ and}$$

$$\Phi(Y) = \begin{bmatrix} \Phi^T(Y_1) \\ \vdots \\ \Phi^T(Y_m) \end{bmatrix} = \eta = \begin{bmatrix} \eta_{11} & \cdots & \eta_{1q} \\ \vdots & \vdots & \vdots \\ \eta_{m1} & \cdots & \eta_{mq} \end{bmatrix}.$$

Define kernel Gram matrices $K_{xx} = \Phi(X)\Phi^T(X)$ and $K_{yy} = \Phi(Y)\Phi^T(Y)$. The dependence measure is

$$D_{YX} = \frac{1}{m^2} \text{Tr}(\hat{K}_{xx}\hat{K}_{yy}) = \frac{1}{m^2} \text{Tr}(\xi\xi^T\eta\eta^T) = \frac{1}{m^2} \text{Tr}(\xi^T\eta\eta^T\xi). \tag{N69}$$

Note that $\xi^T\eta$ asymptotically converges to $E[\xi^T\eta] = \Lambda_{XY}$. Therefore, the dependence measure $D_{YX}$ asymptotically converges to

$$D_{YX} \approx \frac{1}{m^2} \text{Tr}(\Lambda_{XY}\Lambda_{XY}^T) = \frac{1}{m^2} \sum_{i=1}^{p} \sum_{j=1}^{q} \gamma_{ij}^2 = \frac{1}{m^2} r.$$ In other words, the dependence measure is asymptotically equal to the association measure of the FCCA.

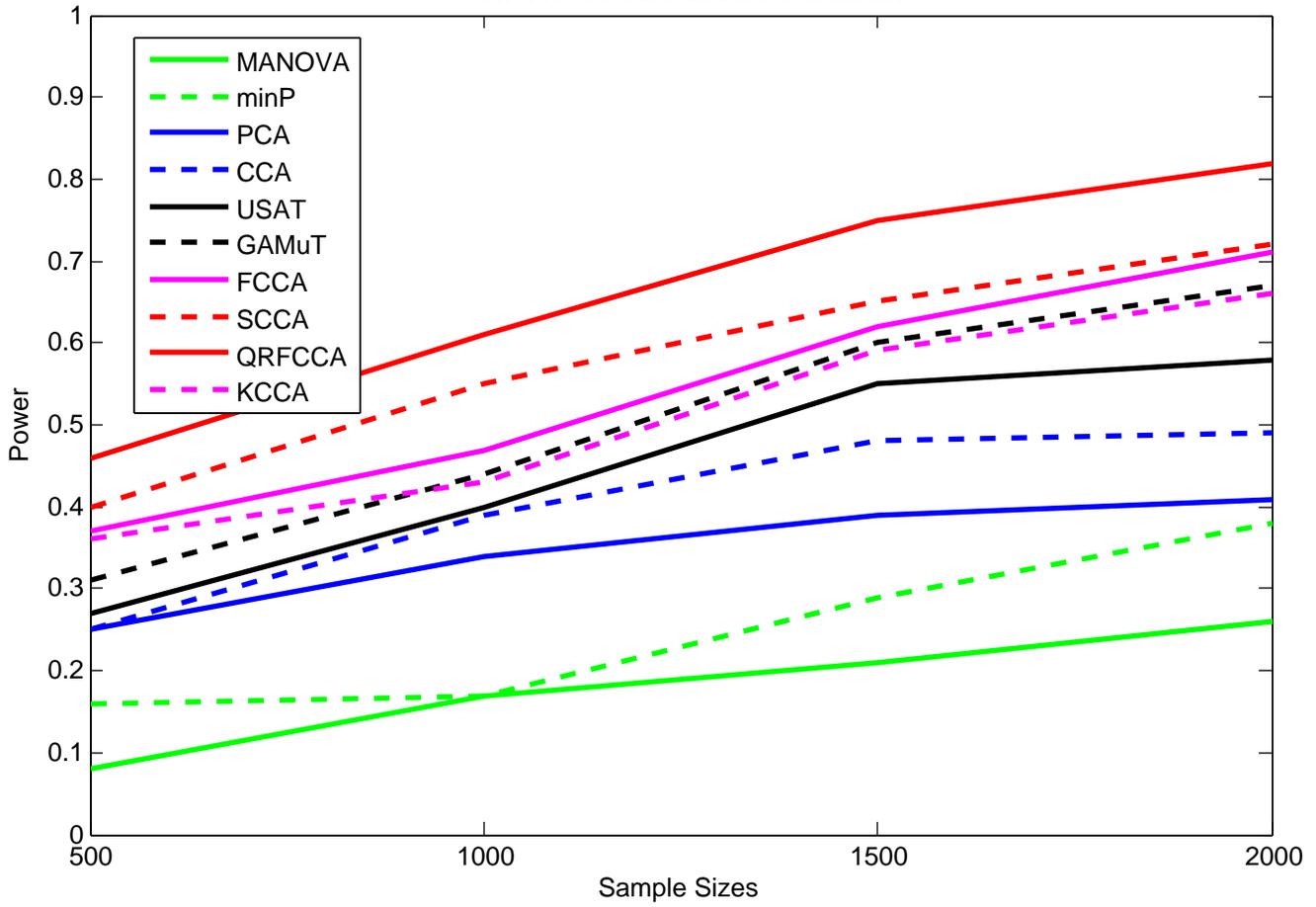

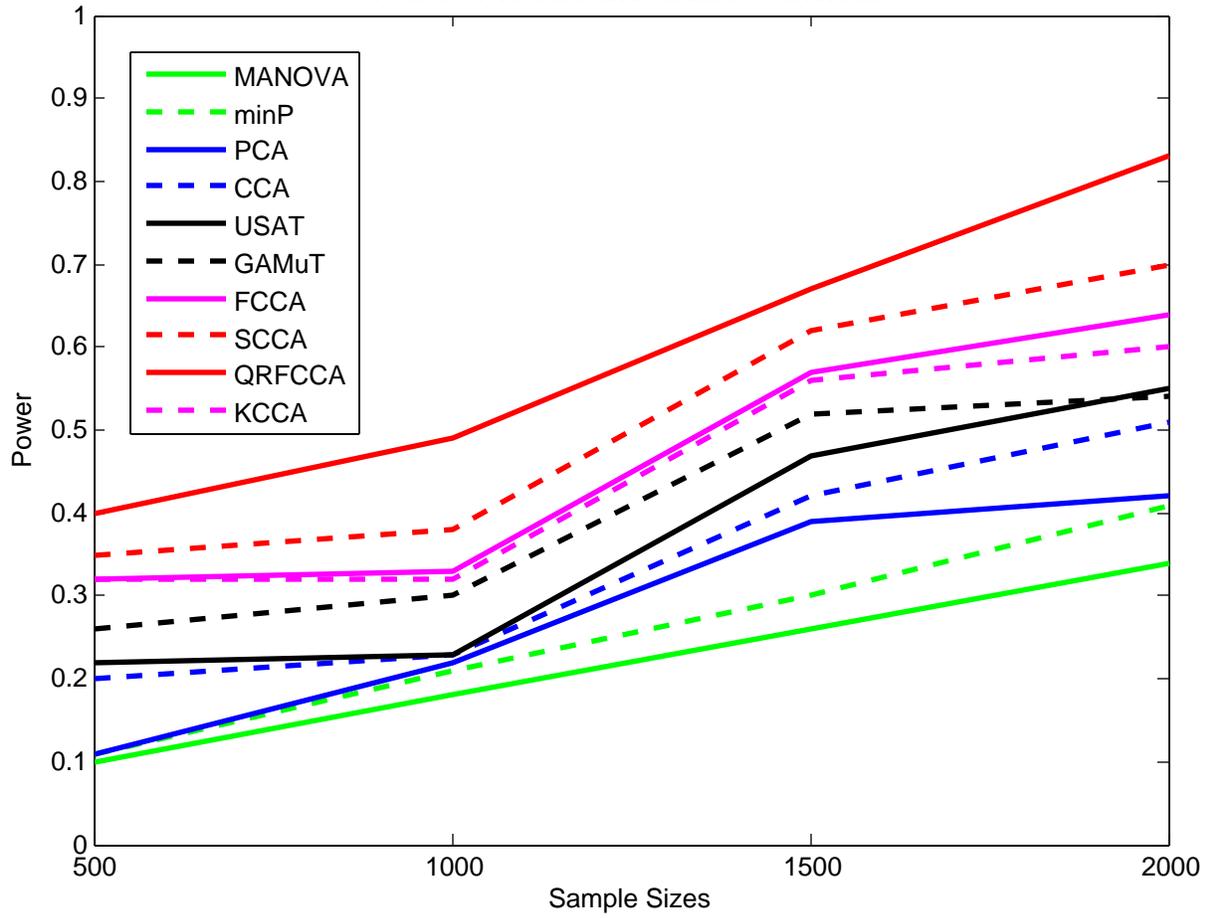

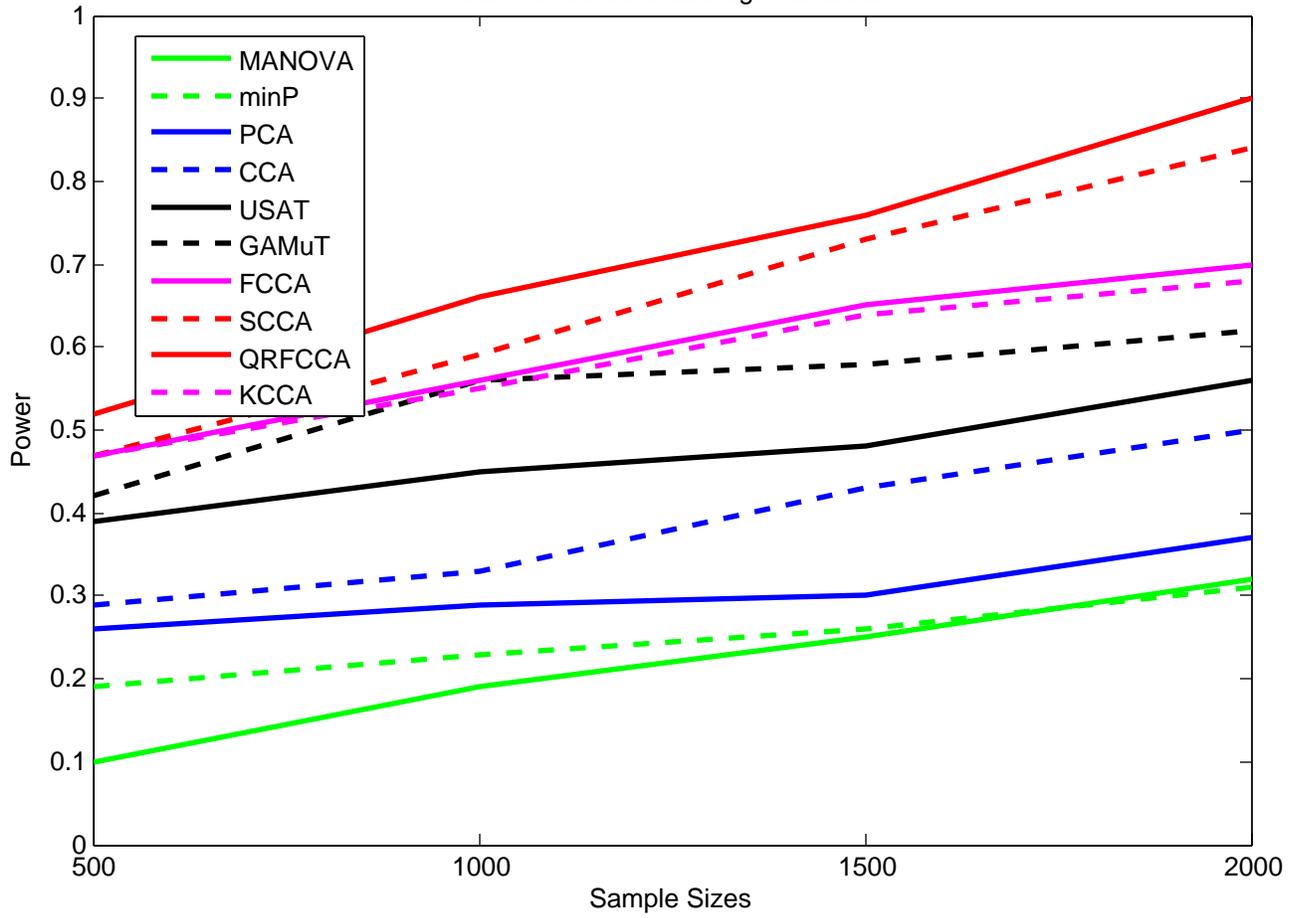

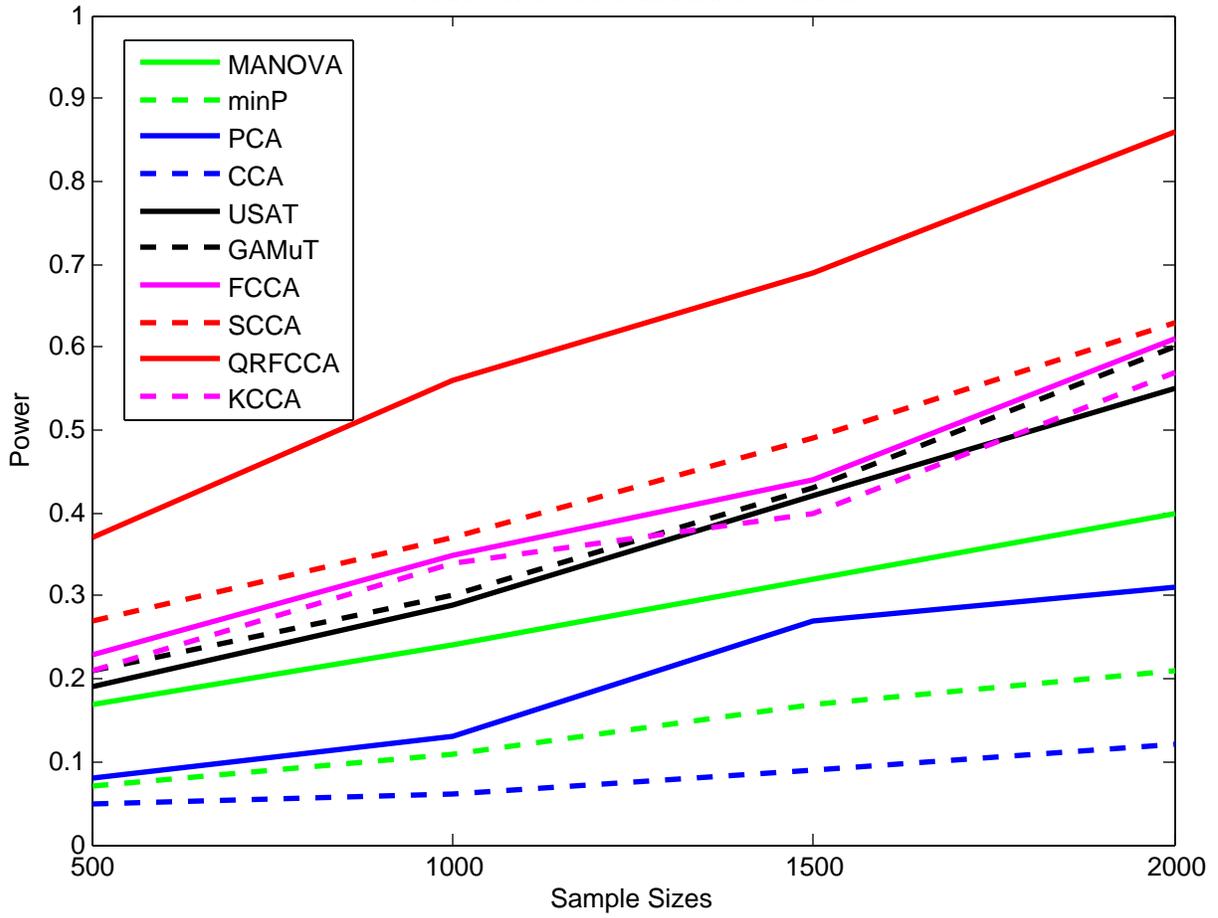

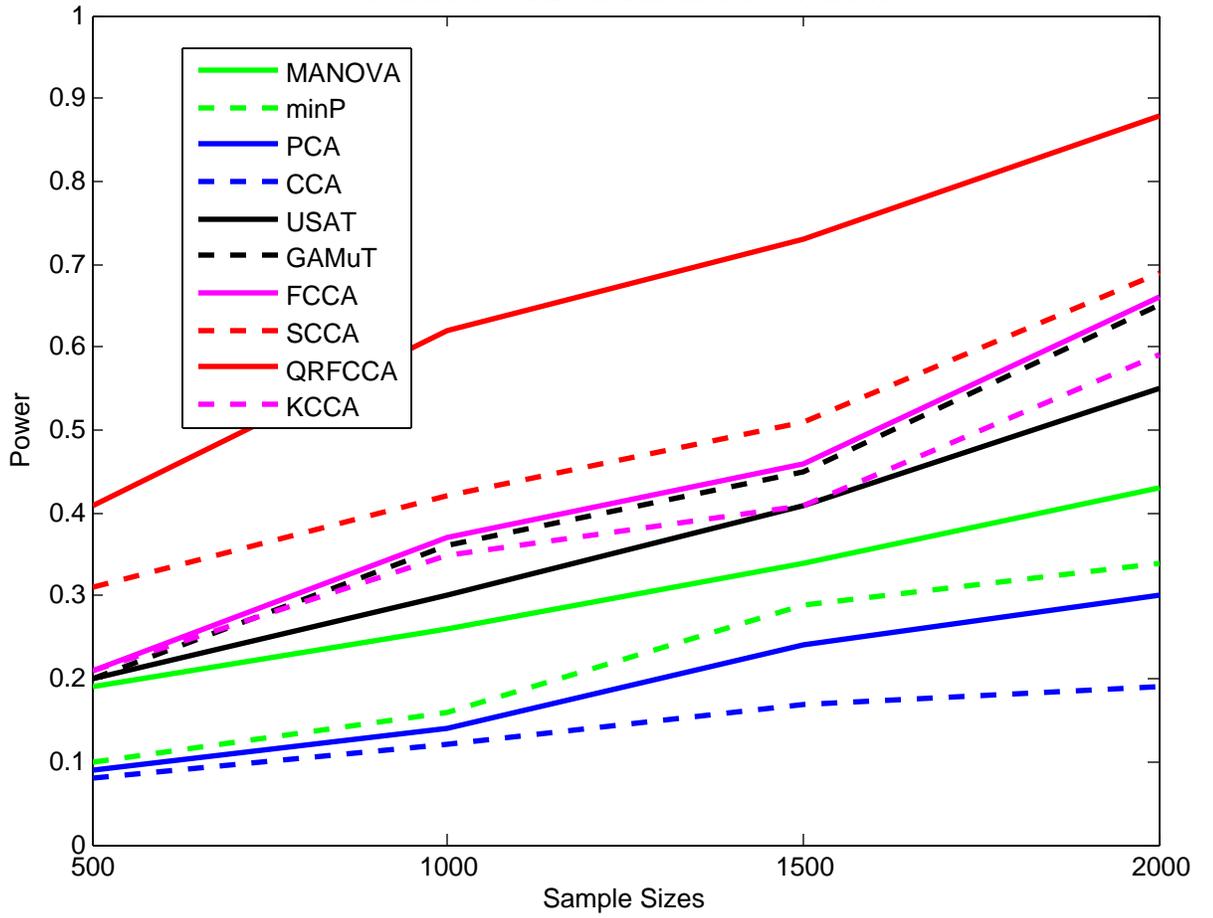

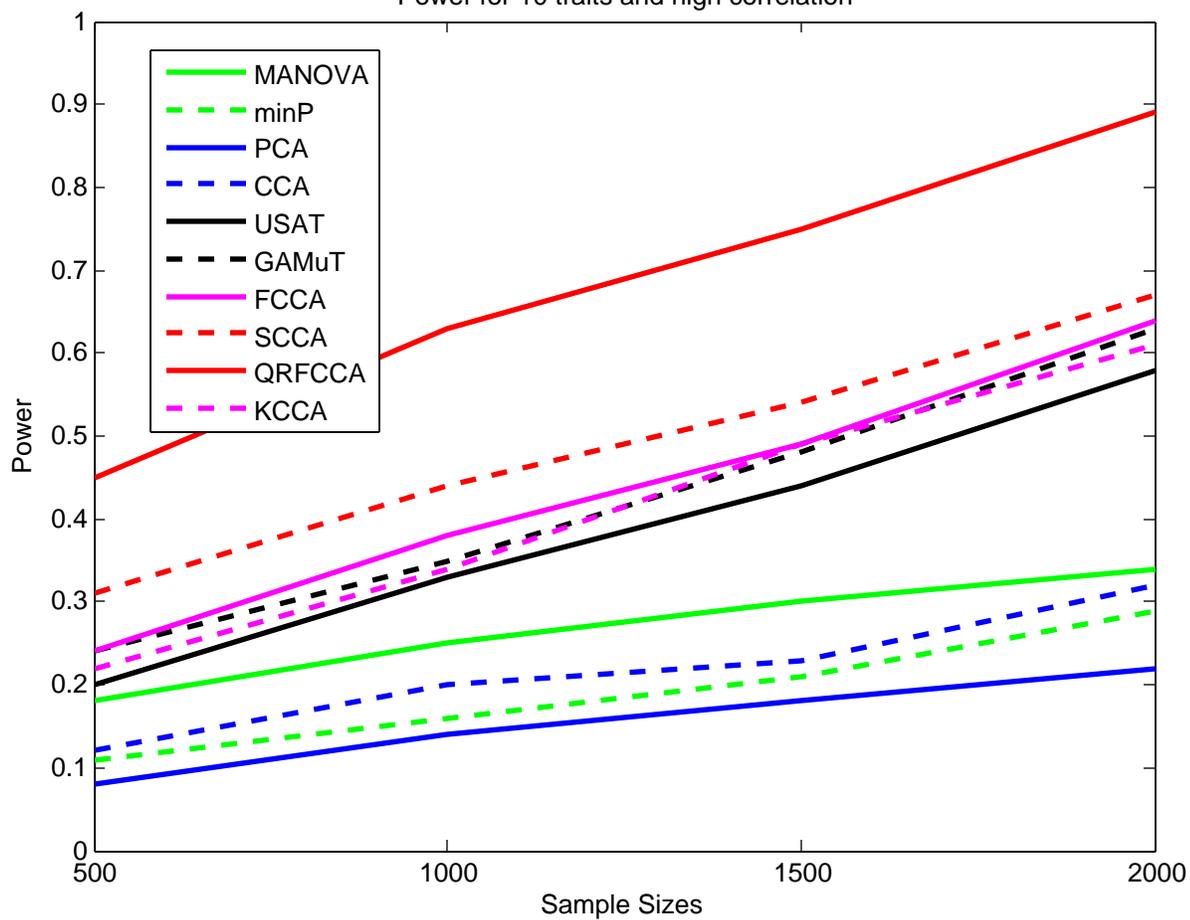

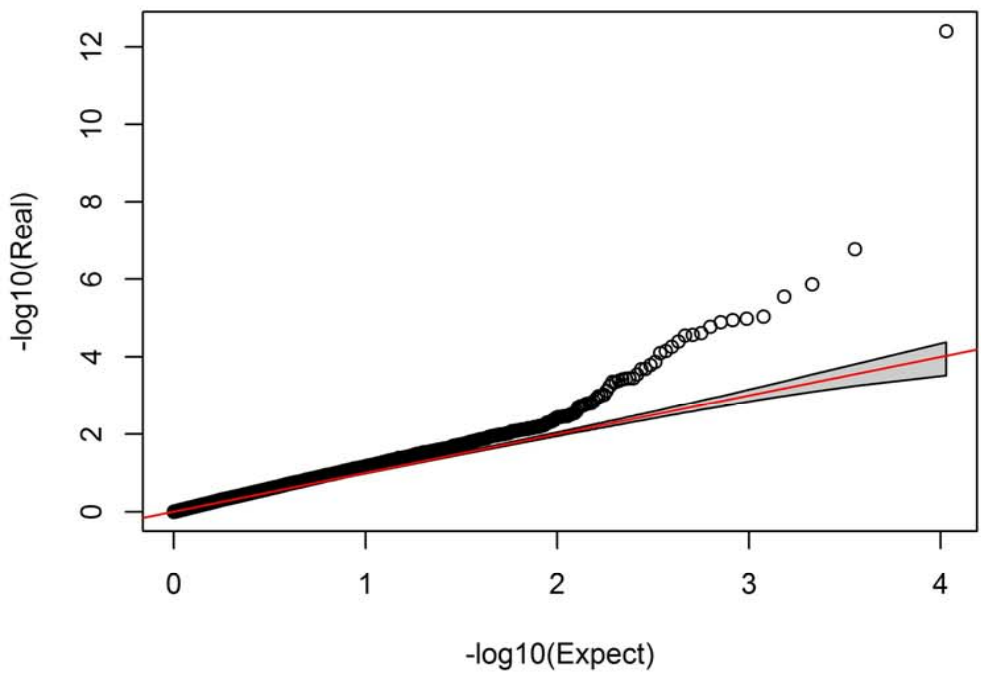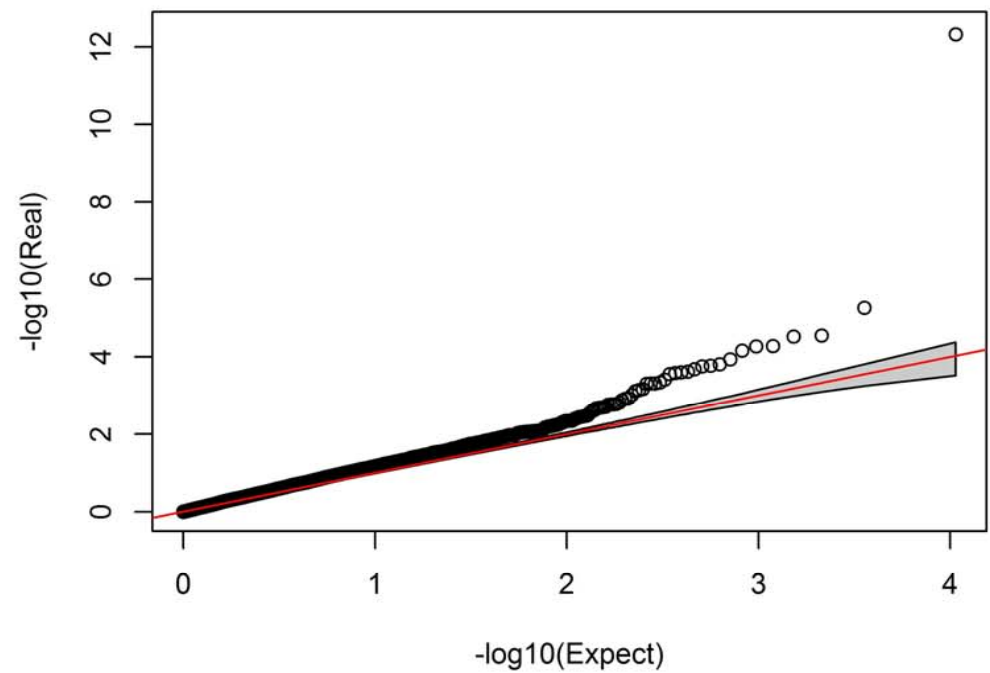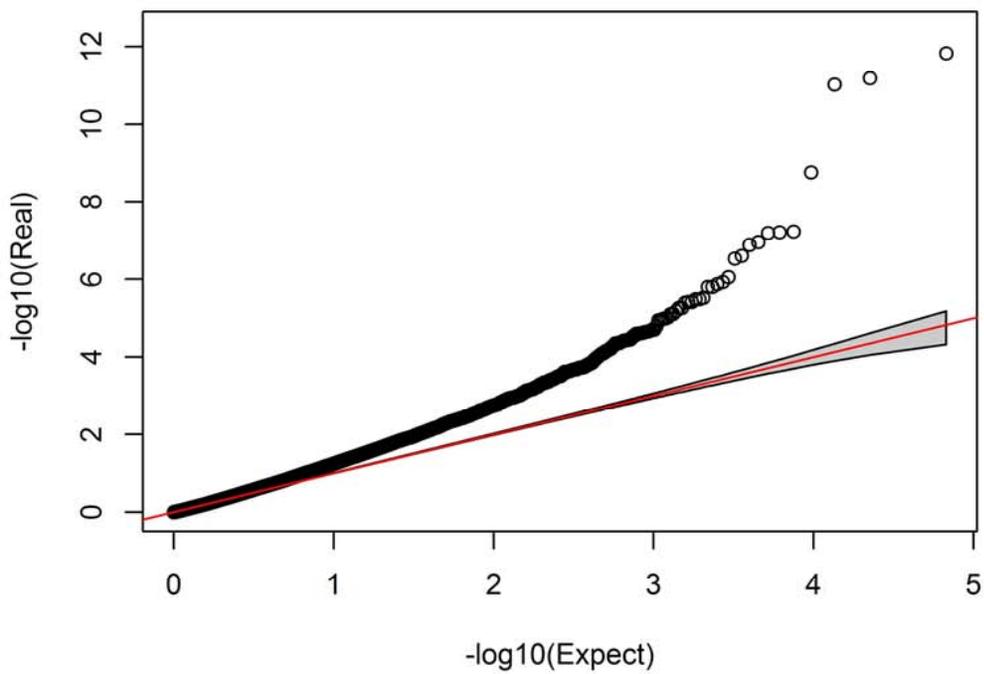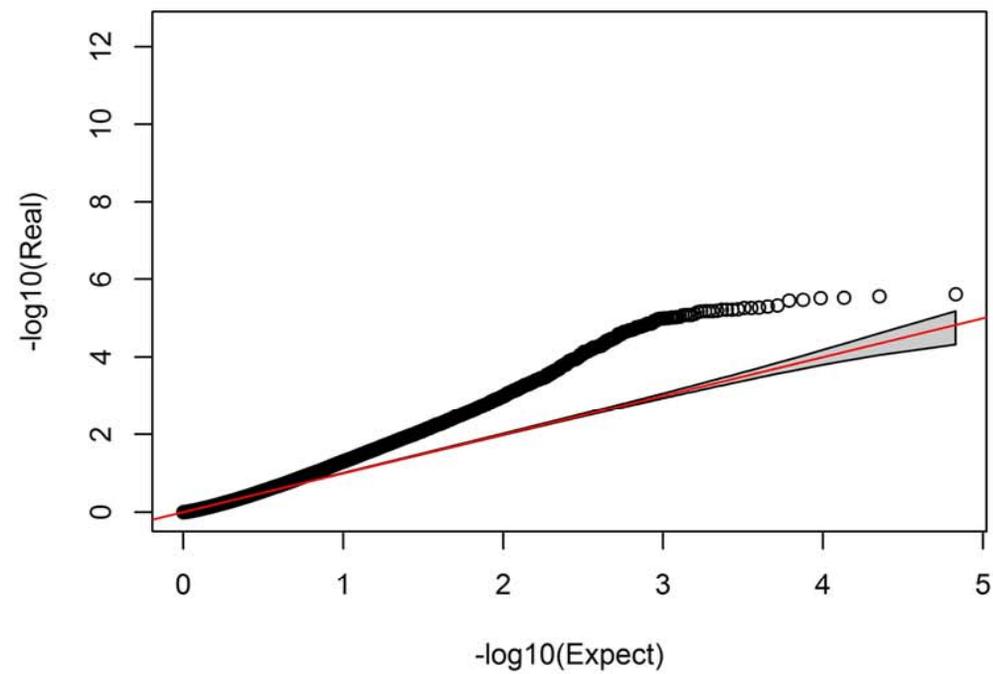

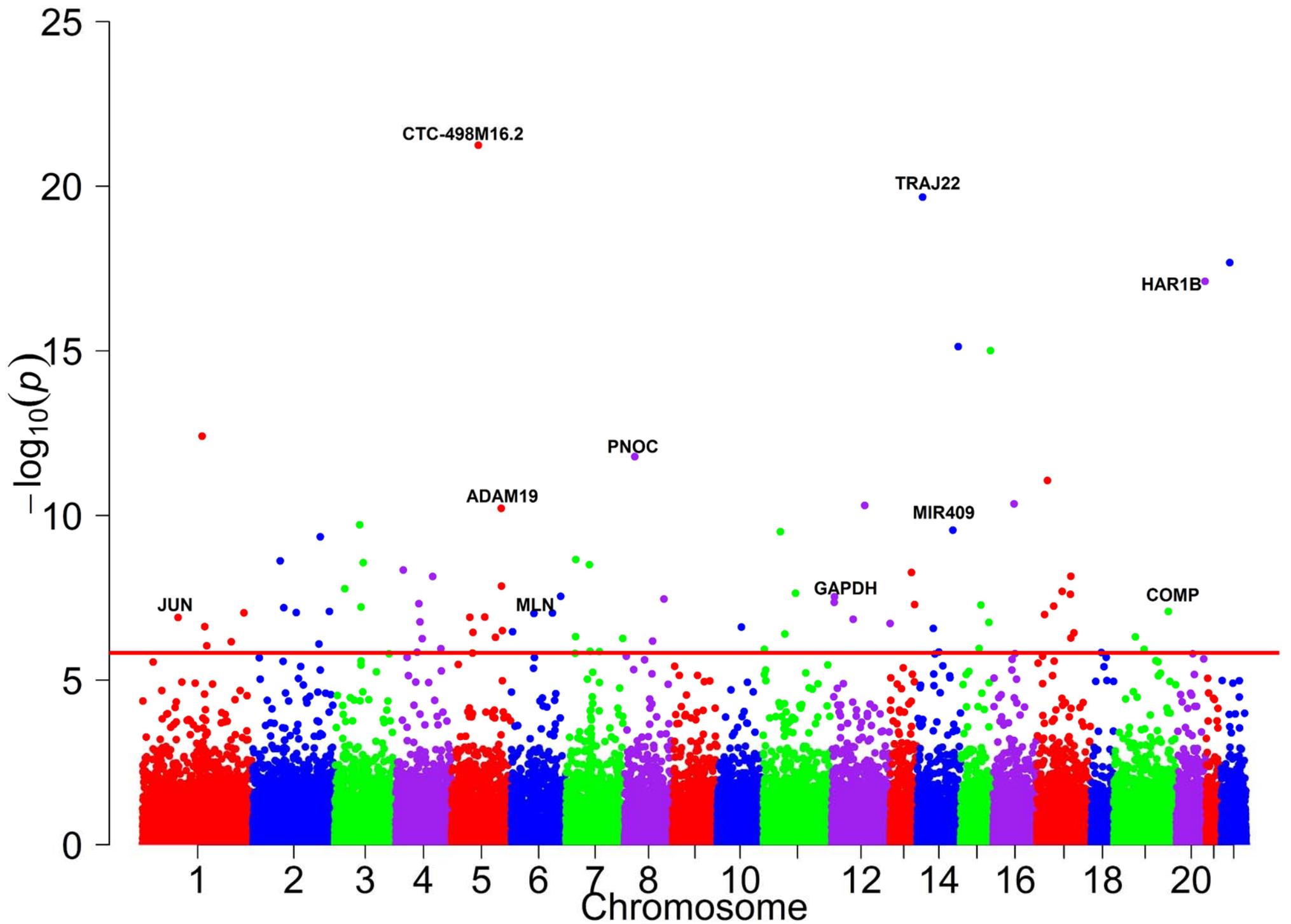

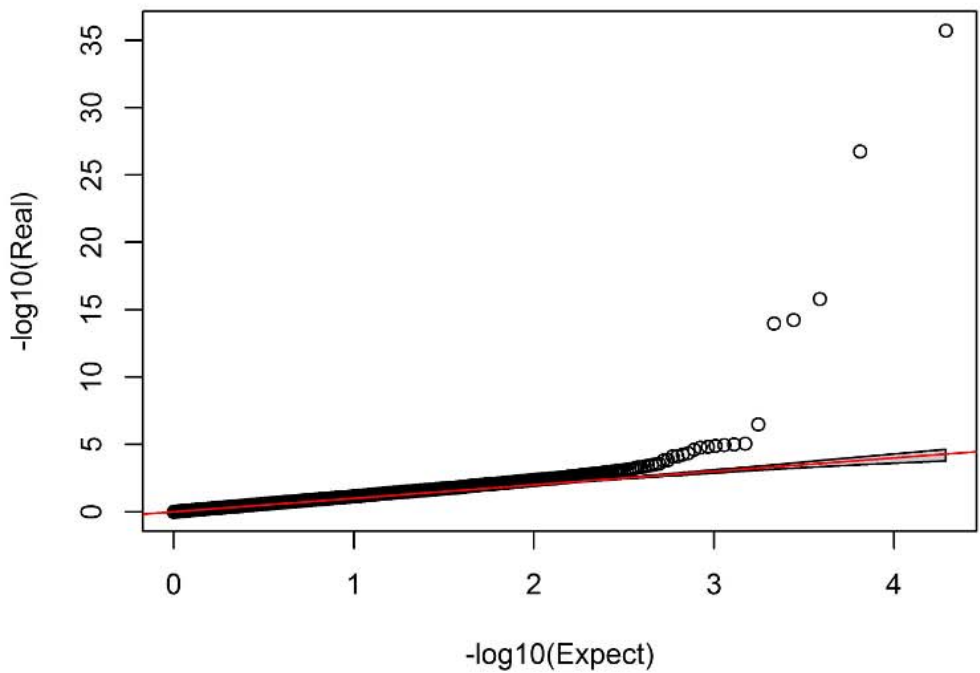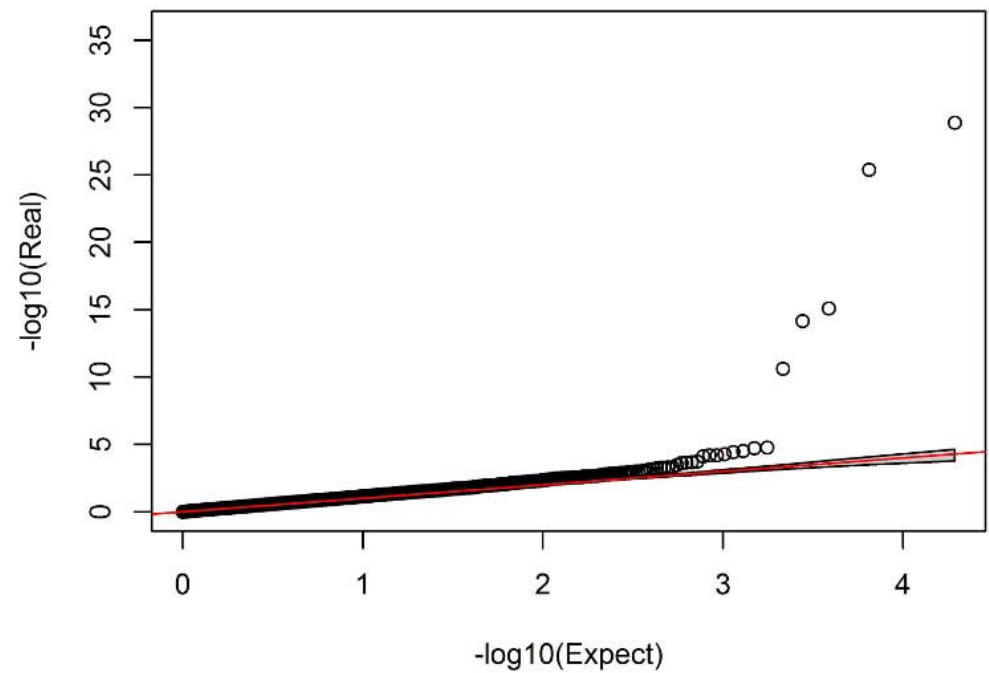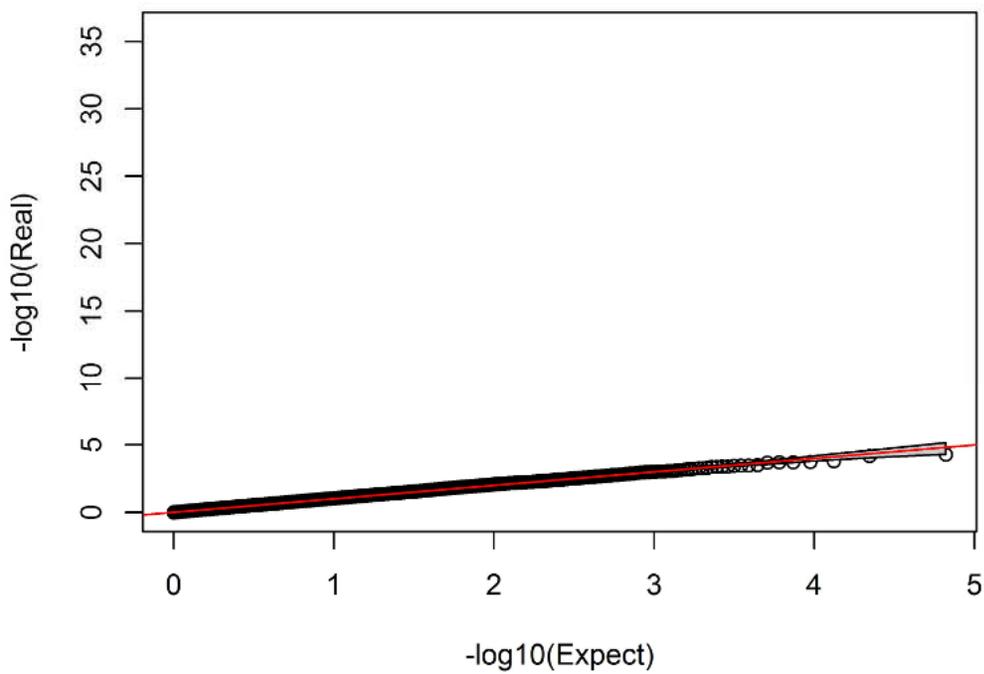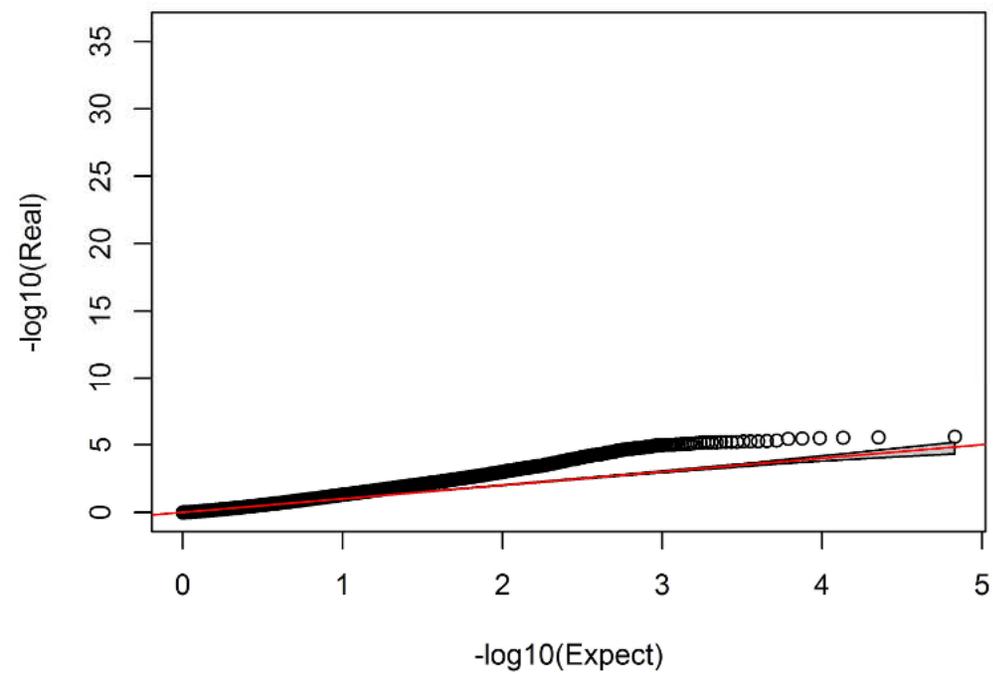

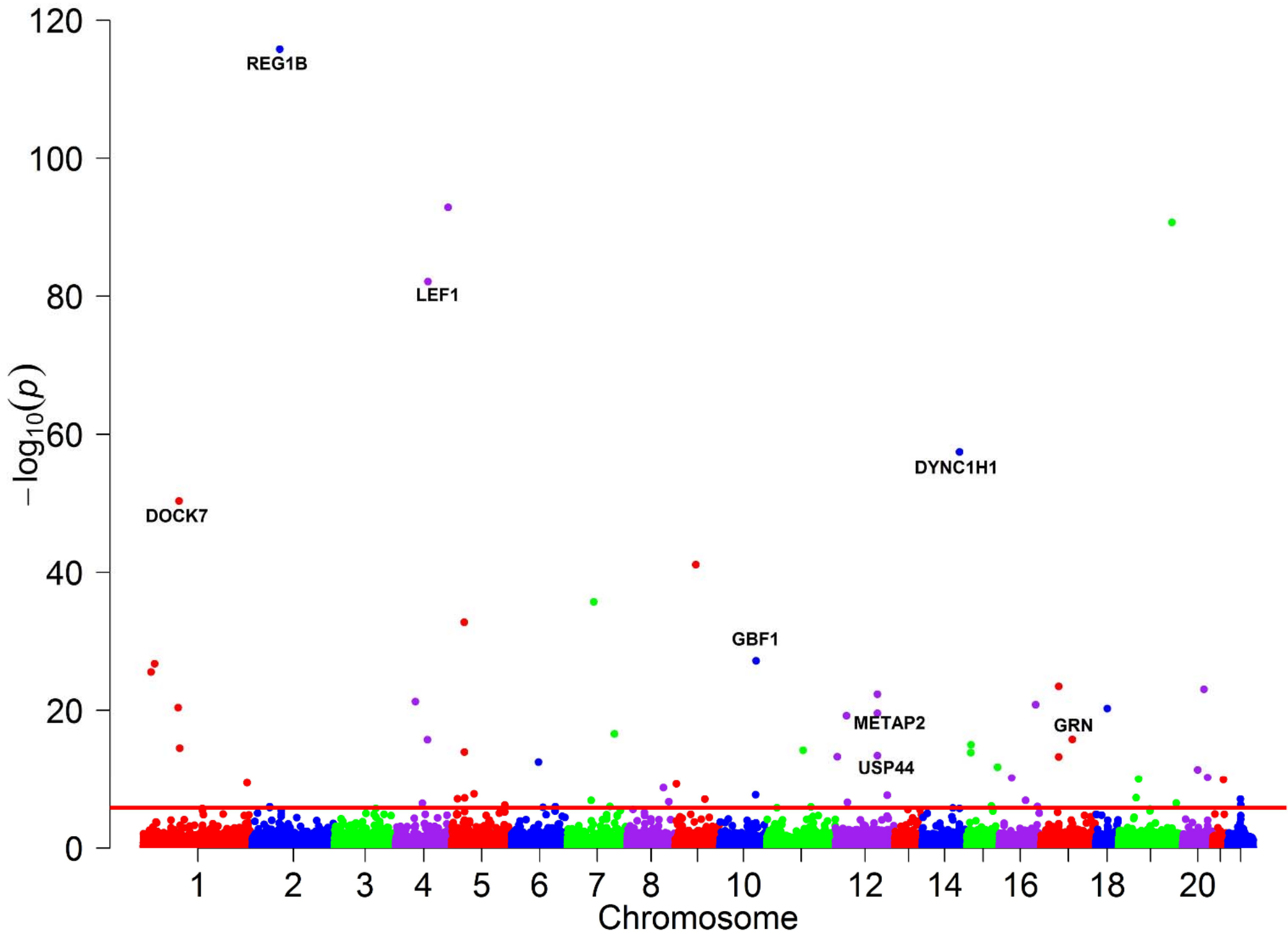

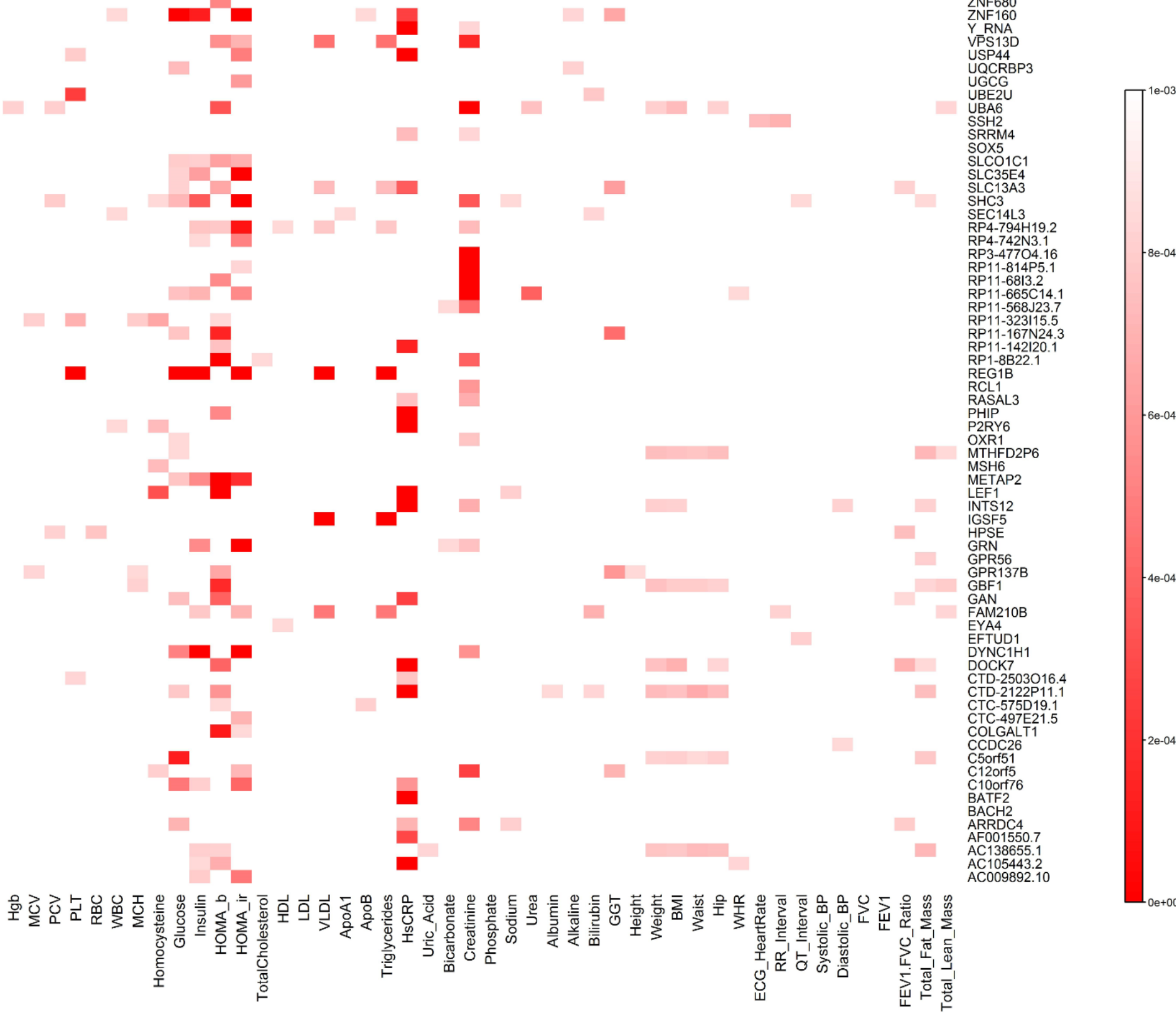